\documentclass[aip,sd,amsmath,amssymb]{revtex4-1}
\usepackage{natbib}
\usepackage{graphicx}% Include figure files
\usepackage{dcolumn}% Align table columns on decimal point
\usepackage{bm}% bold math
\usepackage[utf8x]{inputenc}
\usepackage{color}
\begin{document}

\preprint{AIP/123-QED}

\title{Switching dynamics in cholesteric liquid crystal emulsions}% Force line breaks with \\
%\thanks{Footnote to title of article.}

\author{F. Fadda}
%\email{First.Author@institution.edu.}
\affiliation{Dipartimento di Fisica and Sezione INFN, Università di Bari, Via Amendola 173, 70126 Bari, Italy}

\author{G. Gonnella}
%\email{Third.Author@institution.edu.}
\affiliation{Dipartimento di Fisica and Sezione INFN, Università di Bari, Via Amendola 173, 70126 Bari, Italy}

\author{D. Marenduzzo}
%\email{Fifth.Author@institution.edu.}
\affiliation{SUPA, School of Physics and Astronomy, University of Edinburgh, Edinburgh EH9 3JZ, United Kingdom}%

\author{E. Orlandini}
%\email{Fourth.Author@institution.edu.}
\affiliation{Dipartimento di Fisica e Astronomia  and Sezione INFN, Università di Padova, 35131 Padova, Italy}%

\author{A. Tiribocchi}%
% \email{Second.Author@institution.edu.}
\affiliation{Dipartimento di Fisica e Astronomia and Sezione INFN, Università di Padova, 35131 Padova, Italy}

\date{\today}% It is always \today, today,

\begin{abstract}
In this work we numerically study the switching dynamics of a 2D cholesteric emulsion droplet immersed in an isotropic fluid under an electric field, which is either uniform or rotating with constant speed. The overall dynamics depend strongly on the magnitude and on the direction (with respect to the cholesteric axis) of the applied field, on the anchoring of the director at the droplet surface and on the elasticity. If the surface anchoring is homeotropic and a uniform field is parallel to the cholesteric axis, the director undergoes deep elastic deformations and the droplet typically gets stuck into metastable states which are rich in topological defects. When the surface anchoring is tangential, the effects due to the electric field are overall less dramatic, as a small number of topological defects form at equilibrium. The application of the field perpendicular to the cholesteric axis usually has negligible effects on the defect dynamics. The presence of a rotating electric field of varying frequency fosters the rotation of the defects and of the droplet as well, typically at lower speed than that of the field, due to the inertia of the liquid crystal. If the surface anchoring is homeotropic a periodic motion is found. Our results represent a first step to understand the dynamical response of a cholesteric droplet under an electric field and its possible application in designing novel liquid crystal-based devices.
\end{abstract}

\maketitle

\section{Introduction}\label{Introduction}

Cholesteric liquid crystals are an example of chiral soft material in which the locally favoured state of the director field $\hat{n}$ (a coarse-grain average of the molecular orientation with head-tail symmetry) is a twist deformation in the direction perpendicular to the molecules~\cite{degennes,chandra}. This helical order, usually achieved by adding a chiral dopant in an achiral nematic phase, can be characterized in terms of the helix pitch $p_0$, which is the distance over which the director rotates by $2\pi$. 
When a cholesteric is confined (such as between flat walls), the anchoring on its surface affects the typical helical order and can lead to the emergence of topological defects, whose nature and position crucially depends on the surface anchoring set and on the pitch length~\cite{lavren4,guo}.
Sec et. al.~\cite{zumer}, for instance, have shown that, in a cholesteric droplet with degenerate surface planar anchoring, a proper control of the radius-to-pitch ratio is fundamental to characterize the equilibrium properties of a number of metastable structures as well as the nature of the topological defects. These, in turn, are expected to significantly influence the dynamical response of the director to an external stimulus (such as an electric field or a shear force), as they are the source of flow fields which alter the director orientation\cite{yeomans}. Hence in such systems hydrodynamic interactions are crucial to assess the non-equilibrium physics and simulations are of great importance to capture their role. Indeed, in real experiments the dynamics is often described in terms of coupled differential equations (impossible to threat analytically, unless in the case of weak perturbations), in which the effect of external fields often needs to be taken into account.

Although confined cholesteric liquid crystals have found large technological applications (such as in the design of optical structures used in optics and photonics, as in laser beams~\cite{smal,glees,yang} or in three-dimensional microlasers built from the self-assembly of cholesteric emulsions~\cite{humar}), important questions still need to be addressed. For instance, the application of a strong enough electric field usually causes a profound rearrangement of the director field, often combined with a complex defect dynamics observed in the non-equilibrium regime. The geometry of the confining environment is arguably crucial in determining the out-of-equilibrium physics, as it creates different topological defects at equilibrium that behave differently under an electric field. These also cause the simultaneous presence of regions of different director orientation, which, when subject to the action of an electric field, may align and create complex fluid flow patterns. Such patterns can, in turn, trigger further variation in the director orientation, thereby creating a feedback loop. In addition, the elasticity of the liquid crystal is fundamental as it may create an energy barrier either preventing or favouring (depending on the intensity of the applied field) the formation of metastable states in which the system might get stuck. Even more intriguing is the switching off dynamics, as a proper control of both elasticity and electric field importantly influence the relaxation time towards equilibrium. In general, being able to control the dynamical schedule of the electric field may create versatile routes to create self-assembled materials based on cholesteric emulsions, or on mixtures of cholesterics liquid crystals and isotropic fluids. 

In order to investigate some of these aspects, in this paper we present an initial study about the dynamical response of a droplet of cholesteric liquid crystal in presence of an external electric field. The surface anchoring (either perpendicular or tangential) of the director field and the number of twists of the pitch are key parameters, as they affect both equilibrium properties and dynamical ones. Indeed, at equilibrium several topological defects form, whose nature and position depend on the choice of the anchoring. The presence of an intense enough electric field alter the equilibrium configuration and drives the system towards new metastable states, either quasi-nematic or patterned with several topological defects. The latter, in particular, results when homeotropic anchoring is set, as this accommodates a large number of defects at equilibrium. Switching the field off usually leaves the orientational (director field) profile achieved when the field was on almost unaltered. We have also considered non-uniform fields, such as a time-dependent steadily rotating electric field with a frequency $\omega$. Even in this case the surface anchoring is a key control parameter. In particular, for moderate values of $\omega$ and with homeotropic anchoring we find the droplet undergoes a periodic rotation, whereas the motion becomes irregular for tangential anchoring.

The paper is  organized as follows. In section~\ref{model} we summarize the dynamic equations of motion and the free energy describing a cholesteric droplet hosted in an isotropic fluid. 
In Section~\ref{Equilib} we discuss the equilibrium properties of such system for different anchoring conditions (homeotropic, homogeneous and free) of the director field at the droplet surface and for two helical pitches. 
We next present the main findings of the paper referring to the effect that an external electric field has on the  dynamical properties of the emulsion. Two main protocols are treated. In Section~\ref{switching} the electric field is uniform and, as in common liquid crystals device cells, it is switched ON and OFF to test the response and relaxation properties of the cholesteric droplet as function of the anchoring and helical pitch.
To test the effect that the droplet  curvature may have on the switching properties of the cholesteric emulsion a comparison with the case in which the cholesteric is confined between two parallel walls is also provided. 
In the second protocol, discussed in Section~\ref{rotate}, the electric field is not uniform but rotates continuously with time at a given frequency $\omega$. The interplay between the external electric field and the inertia of the defects dynamics inside the droplet gives rise to interesting non-trivial phenomena, depending on  $\omega$ and on the anchoring conditions.
This is a first step towards a less diluted emulsion  in which many cholesteric droplets share a common region of space.
Section~\ref{conclusions} summarizes the results and concludes the paper.

\section{Model and methods}\label{model}

\subsection{Free energy and equations of motion}
We consider a cholesteric emulsion in a very diluted regime in which a single cholesteric droplet is dispersed in an isotropic fluid. 
In this system the droplet is made up of anisotropic-shaped molecules (such as rods) arranged in a helical fashion, whereas the surrounding phase is modeled as an isotropic liquid crystal in which there is neither orientational nor positional order and molecules are randomly oriented.
The physics of the system is described
in terms of a set of coarse-grained fields, $\rho({\bf r},t)$, $\phi({\bf r},t)$, ${\bf u}({\bf r},t)$
and $Q_{\alpha\beta}({\bf r},t)$ which represent respectively the mass density, a scalar order parameter related to the
concentration of the cholesteric phase relative to the isotropic phase, the average velocity of the fluid and the 
tensor order parameter that, within the Beris-Edwards theory~\cite{beris}, describes the cholesteric phase of the liquid crystal.
In the uniaxial approximation $Q_{\alpha\beta}=q(\hat{n}_{\alpha}\hat{n}_{\beta}-\frac{1}{3}\delta_{\alpha\beta})$ (Greek subscripts denote Cartesian coordinates),
where $\hat{n}$ is the director field which indicates the local average direction of the molecules
and $q$ measures the amount of the local order that is proportional to the largest eigenvalue of ${\bf Q}$ ($0\le q\le\frac{2}{3}$). 
In presence of an external electric field ${\bf E}$ the equilibrium properties of the system  
 are encoded in a Landau-de Gennes free energy ${\cal F} = \int_V f dV$ where 
\begin{eqnarray}\label{free_E}
f&=&\frac{a}{4}\phi^{2}(\phi-\phi_{0})^{2}+\frac{K_{bf}}{2}\left|\nabla \phi\right|^{2}\nonumber\\ &&+
A_0 \biggl[\frac{1}{2}\left(1-\frac{\zeta(\phi)}{3}\right)Q^2_{\alpha\beta}-\frac{\zeta(\phi)}{3}Q_{\alpha\beta}Q_{\beta\gamma}Q_{\gamma\alpha}
                 + \frac{\zeta(\phi)}{4}(Q_{\alpha\beta}^2)^2\biggr]\nonumber\\
       &&+ \frac{K_{lc}}{2}\left[(\partial_{\beta}Q_{\alpha\beta})^2+(\varepsilon_{\alpha\zeta\delta}\partial_{\zeta}Q_{\delta\beta}+2q_0Q_{\alpha\beta})^2\right]\nonumber\\
       &&+W(\partial_{\alpha}\phi)Q_{\alpha\beta}(\partial_{\beta}\phi)-\frac{\epsilon_a}{12\pi}E_{\alpha}Q_{\alpha\beta}E_{\beta}.
\end{eqnarray}

\begin{figure*}[htbp]
\centerline{\includegraphics[width=0.85\textwidth]{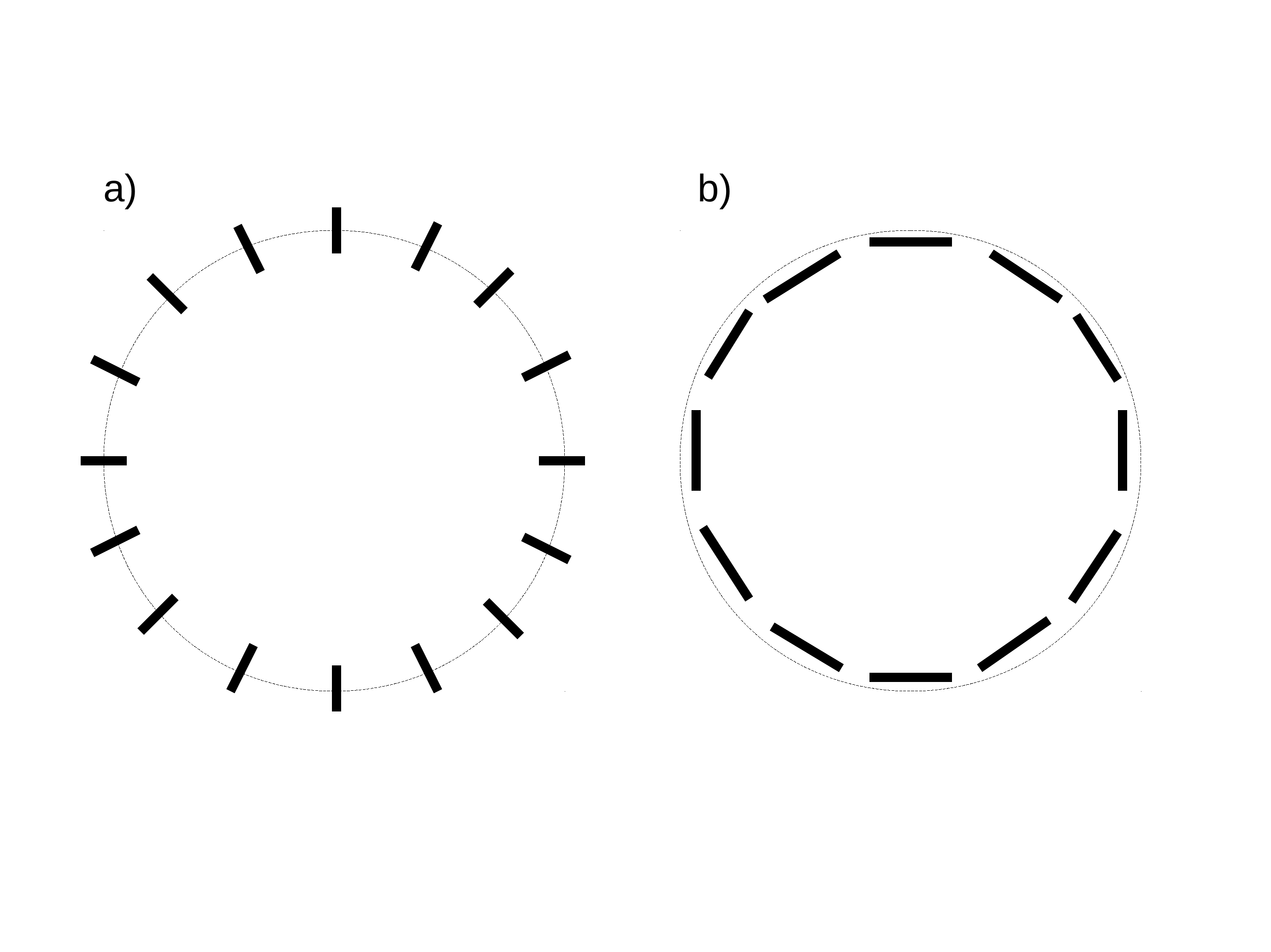}}
\caption{Schematic diagram showing the geometry of the anchoring direction of the director field (thick black lines) at the droplet interface (dashed line). 
a) Homeotropic (perpendicular) anchoring and b) tangential anchoring.}
\label{fig1_N}
\end{figure*}

The first line stems from a typical binary fluid formalism and is made up by
two contributions: the first one, multiplied by the positive constant $a$, is a double well potential
leading to bulk phase separation into a cholesteric (inside the droplet 
with $\phi \simeq \phi_0$) and an isotropic (outside the droplet with $\phi\simeq0$) phase, whereas the second one creates an interfacial tension
between these phases whose strength depends on  $K_{bf}$. Both constants $a$ and $K_{bf}$ also control
the interface width $\Delta$ of the droplet, which goes as $\Delta\sim\sqrt{K_{bf}/a}$.
The second line is made up by three contributions (summation over repeated indexes is assumed) each multiplied by the positive constant $A_0$. 
They represent the  bulk free energy density for an uniaxial liquid crystal 
with an isotropic-nematic  transition at $\zeta(\phi) = \zeta_c=2.7$, where the 
parameter $\zeta(\phi)$ determines which phase is the stable one. If $\zeta(\phi) < \zeta_c$ the phase is isotropic, whereas it is cholesteric if  $\zeta(\phi)>\zeta_c$ (while the chiral terms lead to a deviation of $\zeta_c$ from $2.7$ for the isotropic-nematic transition, this deviation is numerically very small in our case, i.e., $\zeta\simeq 2.7$ in the present case as well).
According to previous studies~\cite{sulaim}, we set $\zeta(\phi)=\zeta_0+\zeta_s\phi$, 
where $\zeta_0$ and $\zeta_s$ are constants controlling the boundary of the coexistence region.
The elastic  penalty due to local  distortions of the cholesteric order
is described by the third line of Eq.(\ref{free_E}) where the (standard) ``one elastic constant'' approximation, $K_{lc}$, 
has been considered~\cite{degennes}.  Notice that here a gradient-independent term of the bulk free energy 
has been incorportated  to have a positive elastic free energy. 
In particular the parameter $q_0=\frac{2\pi}{p_0}$ fixes the pitch length $p_0$ of the cholesteric helix
and $\varepsilon_{\alpha\zeta\delta}$ is the Levi-Civita antisymmetric tensor.
The first  term of the fourth line  takes into accout the anchoring of the  liquid crystal at
the surface of the droplet. The constant $W$ controls the anchoring strength: if negative the director is aligned perpendicularly
(homeotropic anchoring) to the surface, whereas if positive the director is aligned tangentially to the surface (planar anchoring).
In Fig.~\ref{fig1_N} schematic diagrams of these conditions are shown.
Finally the coupling with an external electric field $E_{\alpha}$ is provided by the last term where  the 
dielectric anisotropy $\epsilon_a$  has been fixed to be positive.
The reduced electric  potential is $\Delta V_{y,z}=E_{y,z} L_{y,z}$, where $E_{y,z}$ is the component of the electric field 
along the $y,z$-axis and $L_{y,z}$ is the size of the cell.
In presence of confining walls, one has to take into account also the energy due to the 
anchoring of the liquid crystal with these boundaries.
This is described by a surface term 
\begin{equation}
\frac{1}{2}\alpha_s(Q_{\alpha\beta}-Q^0_{\alpha\beta})^2,
\end{equation}
where the constant $\alpha_s$ controls the strength of the pinning at the walls of the director
and $Q^0_{\alpha\beta}=S_0(n^0_{\alpha}n^0_{\beta}-\delta_{\alpha\beta}/3)$, with $S_0$ and 
$n^0_{\alpha}$ being respectively the magnitude and the direction  of the cholesteric ordering at the walls.

It is customary to write the free energy in terms of a minimal set of dimensionless parameters  which identify the 
position of the system in the thermodynamic phase space~\cite{mermin}.  These are
\begin{eqnarray}
\kappa&=&\sqrt{\frac{108K_{lc}q_0^2}{A_0\zeta(\phi_0)}},\\
\tau&=&\frac{27(1-\zeta(\phi_0)/3)}{\zeta(\phi_0)},\\
{\cal E}^2&=&\frac{27\epsilon_a}{32\pi A_0\zeta(\phi_0)}E_{\alpha}E_{\alpha},
\end{eqnarray}
where $\tau$ is the reduced temperature, which multiplies the quadratic terms of  the dimensionless bulk free energy, and $\kappa$ is the chirality which multiplies the gradient
free energy terms. This parameter measures the amount of twist stored in the system  and is zero when $q_0=0$. This means that 
the corresponding phase can be either nematic or isotropic.  In this case to unambiguously determine the phase of the system
the value of $\zeta(\phi)$ must be given.

The dynamical equations governing the evolution of the system are

\begin{eqnarray}
\partial_{t}\phi+ \partial_{\alpha}(\phi u_{\alpha})&=&\nabla \left (M \nabla\mu\right),\label{phi_eq}\\
(\partial_t+{\bf u}\cdot\nabla){\bf Q} - {\bf S}({\bf W},{\bf Q})&=&\Gamma {\bf H},\label{Q_eq}\\
\nabla\cdot{\bf u}&=&0,\label{cont}\\
\rho(\partial_t+u_{\beta}\partial_{\beta})u_{\alpha}&=&\partial_{\beta}\sigma_{\alpha\beta}^{total}.\label{nav_stok}
\end{eqnarray}
Eq.(\ref{phi_eq}), that  describes the evolution of the concentration $\phi({\bf r},t)$, is a convection-diffusion equation 
of a model B~\cite{bray}, where $M$ is the mobility and $\mu = \delta {\cal F} / \delta \phi$ is the chemical potential. 
The dynamics of the liquid crystal order parameter ${\bf Q}$ is governed
by the convection relaxation equation~(\ref{Q_eq}). 
The first two terms on the left hand side are the
material derivative. Moreover, since for rod-like molecules
the order parameter distribution can be rotated and stretchedy
by flow gradients, a further contribution ${\bf S}({\bf W},{\bf Q})$  is needed.
This is given by~\cite{beris}
\begin{equation}\label{eq_S}
{\bf S}({\bf W},{\bf Q})=(\xi{\bf D}+{\bf \Omega})({\bf Q}+{\bf I}/3)+({\bf Q}+{\bf I}/3)(\xi{\bf D}-{\bf \Omega})
-2\xi({\bf Q}+{\bf I}/3)Tr({\bf Q}{\bf W}),
\end{equation}
where ${\bf D}=({\bf W}+{\bf W}^T)/2$ and ${\bf \Omega}=({\bf W}-{\bf W}^T)/2$ are the symmetric and antisymmetric parts
of the velocity gradient tensor $W_{\alpha\beta}=\partial_{\beta}u_{\alpha}$ and ${\bf I}$ is the unit matrix. 
The constant $\xi$ plays a double role:  it controls the dynamical response of the liquid crystal when a shear flow is applied, 
in particular whether it is either flow tumbling ($\xi < 0.6$) or flow aligning ($\xi > 0.6$); it also defines the aspect 
ratio of the molecules of the liquid crystal, positive for rod-like molecules and negative for the oblate ones. 
In all the simulations considered we have taken $\xi=0.7$ that desribes  rod-like molecules in flow-aligning regime.
In the right-hand side of Eq.(\ref{Q_eq}) $\Gamma$ is the collective rotational diffusion constant while 
the molecular field ${\bf H}$ is given by
\begin{equation}
{\bf H}=-\frac{\delta {\cal F}}{\delta {\bf Q}}+\frac{{\bf I}}{3}Tr\frac{\delta {\cal F}}{\delta {\bf Q}}.
\label{H_eq}
\end{equation}
The last two equations are the continuity and the Navier-Stokes equations in the limit of an incompressible fluid 
and $\sigma^{total}$ is the total hydrodynamic stress given by
\begin{equation}
\sigma^{total}_{\alpha\beta}=\sigma^{visc}_{\alpha\beta}+\sigma^{el}_{\alpha\beta}+\sigma^{int}_{\alpha\beta},
\end{equation}
where $\sigma^{visc}$, $\sigma^{el}$ and $\sigma^{int}$ are, respectively, the viscous stress, the stress due 
to the liquid crystalline order and the interfacial stress between the isotropic and the cholesteric phase. 
Their explicit expressions are~\cite{beris,degennes}
\begin{eqnarray}
\sigma^{visc}_{\alpha\beta}&=&\eta(\partial_{\alpha}u_{\beta}+\partial_{\beta}u_{\alpha}),\\
\sigma^{el}_{\alpha\beta}&=&-P\delta_{\alpha\beta}-\xi H_{\alpha\gamma}(Q_{\gamma\beta}+\frac{1}{3}\delta_{\gamma\beta})-\xi(Q_{\alpha\gamma}+\frac{1}{3}\delta_{\alpha\gamma})H_{\gamma\beta}\nonumber\\&&+
2\xi(Q_{\alpha\beta}-\frac{1}{3}\delta_{\alpha\beta})Q_{\gamma\mu}H_{\gamma\mu}+Q_{\alpha\nu}H_{\nu\beta}-H_{\alpha\nu}Q_{\nu\beta},\label{el_sigma}\\
\sigma_{\alpha\beta}^{int}&=&-\left(\frac{\delta{\cal F}}{\delta\phi}\phi-{\cal F}\right)\delta_{\alpha\beta}-\frac{\delta{\cal F}}{\delta(\partial_{\beta}\phi)}\partial_{\alpha}\phi-
\frac{\delta{\cal F}}{\delta(\partial_{\beta}Q_{\gamma\mu})}\partial_{\alpha}Q_{\gamma\mu},
\end{eqnarray}
where $\eta$ is the isotropic shear viscosity and $P$ is the isotropic pressure. The complex expression of the total stress tensor reflects the additional complications of liquid crystal hydrodynamics, in which an antisymmetric contribution, given by $Q_{\alpha\nu}H_{\nu\beta}-H_{\alpha\nu}Q_{\nu\beta}$
(appearing in Eq.~\ref{el_sigma}) is necessary alongside the symmetric part (all the remaining terms). If more elastic constants were considered, further antisymmetric terms would come from
$\frac{\delta{\cal F}}{\delta(\partial_{\beta}Q_{\gamma\mu})}\partial_{\alpha}Q_{\gamma\mu}$.

We note that, while the form of our stress tensor is not the most general one provided in Ref.~\cite{beris}, our equations account for the main phenomenology of liquid crystal hydrodynamics~\cite{Denniston}. Our equations can be mapped onto the Leslie-Ericksen (LE) theory in the uniaxial limit, and in particular the six Leslie coefficients $\alpha_i$ ($i=1,\ldots,6$)~\cite{degennes,Ericksen,Leslie} can be explicitly written as a function of the parameters in our theory~\cite{beris,Grmela1990,Denniston}. The nematodynamic equations we consider are more general than the LE theory because they can describe the hydrodynamics of topological defects, that are characterised by an abrubt decrease of the magnitude of orientational order in their neighborhoods~\cite{Denniston}.

\subsection{Numerical aspects and mapping to physical units}\label{numerics}

Eqs.(\ref{phi_eq},\ref{Q_eq},\ref{cont},\ref{nav_stok}) are numerically integrated by using a scheme in which a standard lattice Boltzmann approach~\cite{soft_cates} is adopted to solve the Navier-Stokes equation while a finite difference scheme, based on a predictor-corrector algorithm, was implemented to solve Eq.(\ref{phi_eq}) and Eq.(\ref{Q_eq}). This hybrid integration scheme was successfully adopted in previous studies on phase separating binary fluids~\cite{tiribocchi}, on active fluids~\cite{marenduzzo,nat_comm}, on cholesterics~\cite{orlandini,orlandini2,lintuvuori,lintuvuori2} 
and on blue phases~\cite{soft_cates,soft,prl}.
Most of the simulations were performed on a quasi-2D lattice ($L_y=128$, $L_z=128$), in which the macroscopic variables are defined on the $yz$ plane. In contrast to a ``fully 2D'' geometry, in which the director and velocity fields are constrained to lie on the $yz$ plane, in the ``quasi-2D'' case an out-of-plane component along the $x$-direction is allowed. This construction is used to fully accommodate the inherent three-dimensional structure of the cholesteric helix.

The system is initialized by setting the concentration field $\phi$ and the order parameter ${\bf Q}$ equal to zero outside the droplet,  to describe an isotropic phase. Inside the droplet $\phi$ is fixed to a constant value $\phi_0$ while ${\bf Q}$ is chosen  to accommodate a cholesteric liquid crystal with helical axis parallel to the $y$-axis (see Fig.~\ref{fig1}). This is ensured by setting
\begin{eqnarray}
Q_{xx}&=& (c_0-c_1/2)\cos(2q_0y)  + c_1/2,\\
Q_{yy}&=& -c_1,\\
Q_{xz}&=& -(c_0-c_1/2)\sin(2q_0y),
\end{eqnarray}
where $c_0=0.546$, $c_1=0.272$, while $Q_{xy}=Q_{yz}=0$. 
The number of twists of the helix is controlled by the parameter $q_0=2\pi/p_0$ through the pitch length $p_0$. In order to compare the cholesteric pitch with the size of the droplet, it is customary to define the pitch length in terms of the droplet diameter, $2R$, as $p_0=4R/N$. In other words  $N$ is the number of $\pi$ twists the cholesteric liquid crystal would display over a distance of $2R$ if not confined. 

Finally at the droplet surface we impose either homogeneous (tangential) or  homeotropic (perpendicular) strong anchoring of the director field, by setting the anchoring strength $W=\pm 0.04$.
The system is then let to relax into its equilibrium state, that is achieved when the total free energy of the system reaches a constant minimum value. 
Afterwards an electric field, either uniform or rotating, is switched on and the time evolution of the system is monitored until it reaches a steady state (ON). A relaxation dynamics toward an OFF state is also studied.

In most of the simulations the following parameter values have been used: $a= 0.07$, $K_{bf}=0.14$, $M = 0.05$, $\phi_0 =2$, $\Gamma = 1$, $A_0 = 1$ and $\eta=1.67$. 
We have also fixed $\tau=0$ (namely $\zeta(\phi_{0})=3$) and tuned the parameters $K_{lc}$ and $q_0$ (which control the chirality $\kappa$)  to set the phase inside the droplet to be a 
cholesteric one~\cite{dupuis,henrich}. In particular $K_{lc}$ ranges from $0.03$ to $0.001$ and $q_0=2\pi/32, 2\pi/64$. 
Hence if $A_0 = 1$, $K_{lc}=0.03$, $\zeta(\phi_{0})=3$ and $q_0=2\pi/32$, one can describe a cholesteric phase whose chirality $\kappa \sim 0.2$, with a rotational viscosity of $1$ Poise, a Frank elastic constant of roughly $3\times 10^{-11}$ N, and a droplet size of $1-10 \mu$m. This corresponds to a length-scale of $\Delta x=10^{-7}$m and a time-scale of $\Delta t=10^{-6}$s, where $\Delta x$ and $\Delta t$ are the lattice-step and the time-step in our simulations. Furthermore the mapping with the electric field can be made by using the previously defined dimensionless number ${\cal E}$. An electric field of $1-10$ $V/\mu$m and a dielectric anisotropy of order $10$, by assuming $\frac{27}{2 A_0\zeta(\phi_{0})}\simeq 2-5\times 10^{-6} J^{-1}m^3$(see Ref.~\cite{mermin}), gives ${\cal E}^2\sim 0.0036$. Lastly, our choice of strong surface anchoring corresponds to an experimental value of $W\sim 10^{-6} J m^{-2}$~\cite{sulaim,poon}.

\section{Equilibrium properties of a cholesteric droplet}\label{Equilib}
In this section we present the results on a droplet of cholesteric liquid crystal equilibrated in an isotropic fluid. We will focus in particular on the impact that the anchoring strength and direction of the cholesteric at the droplet surface have on the equilibrium configurations.  

We consider the case of a cholesteric droplet of radius $R=32$ lattice units initialised at the center of a simulation box of size $1\times 128\times 128$.  Periodic boundary conditions are imposed along all directions; hence the problem is two dimensional. 
Each simulation is run for $4\times 10^5$ time-steps, until the droplet is completely equilibrated. This is determined by looking at the relaxation  behavior of the free energy towards a constant minimum value.

\begin{figure*}[htbp]
\centerline{\includegraphics[width=0.72\textwidth]{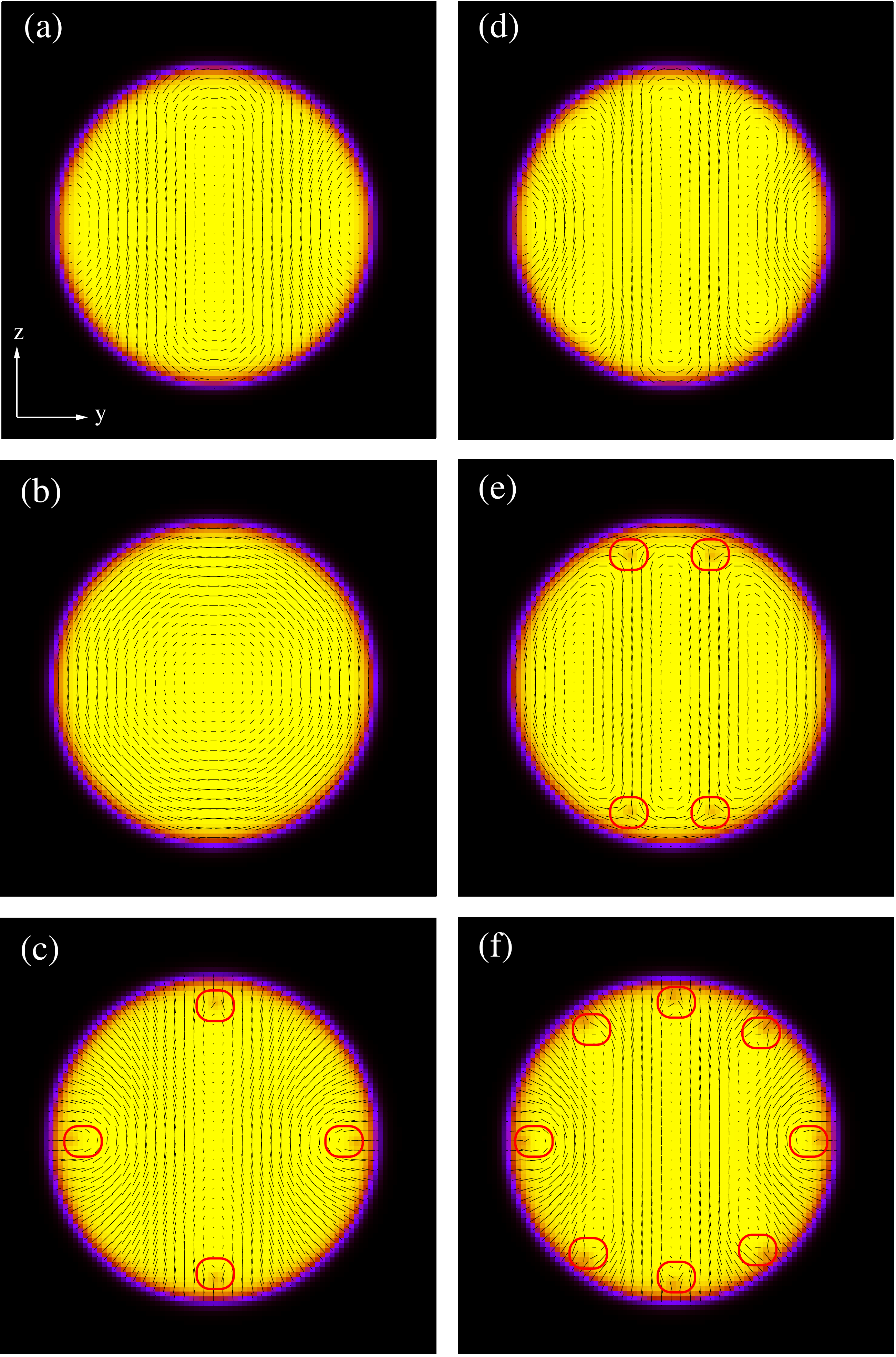}}
\caption{Equilibrium configurations (taken at simulation time $t=4\times 10^5$) of a cholesteric droplet in an isotropic fluid. The radius of the droplet is $R=32$ lattice units and the cholesteric pitch is $q_0=2\pi/64$ ((a)-(b)-(c)) and $q_0=2\pi/32$ ((d)-(e)-(f)). In (a) and (d) no anchoring is set, in (b) and (e) strong tangential anchoring is set while in (c) and (f) strong homeotropic anchoring is chosen. Topological defects, formed nearby the surface (namely $\tau$ and twist disclinations), are highlighted with red circles. The color map represents the largest eigenvalue of the ${\bf Q}$ tensor and ranges from 0 (the black region outside the droplet) to $\simeq 0.33$ (the yellow region inside the droplet). The color map applies to all figures.}
\label{fig1}
\end{figure*}

In Fig.~\ref{fig1} we show a set of equilibrium configurations obtained in absence of surface anchoring (top row) and by imposing either tangential anchoring (middle row) or homeotropic anchoring 
(bottom row). The last two anchoring conditions are very much used in experiments as they are relatively easy to realize and control. Nowadays there are indeed a number of well established techniques 
to accurately control and manipulate the average alignment direction and  magnitude of the anisotropic molecules
 at a droplet interface~\cite{poulin} as well as at a colloid surface~\cite{hijnen,lintuvuori2} 
or at the boundary of a liquid crystalline device~\cite{yokoyama}.
We have generally looked at two cases: the first one with $q_0=2\pi/64$ giving a helical phase with $N=2$ twists (left column)  and the second one obtained with $q_0=2\pi/32$ and hence a helical phase with
$N=4$ twists (right column, see caption of Fig.~\ref{fig1} for more details).
The figure shows a clear dependence of the equilibrium configuration both on the anchoring  and on the number of twists accommodated inside the droplet. As expected, for very weak anchoring the director field orients itself continuously at the boundary to satisfy the natural curvature of the droplet. Hence the resulting equilibrium configurations are defect-free (see Fig.~\ref{fig1}a,d). 
For strong anchoring, frustration between the cholesteric order inside the droplet and the imposed direction at the surface is present and  equilibrium configurations with defects are expected~\cite{sulaim,tiribocchi2,zumer}. 
Defects can result either from sharp changes in orientation of the director in the interior of the droplet, or due to conflict with the anchoring close to the boundary. 
In addition we find that their number and position depend on the anchoring direction and strength, as well as on the number $N$ of cholesteric twists in the droplet. Defects in cholesterics are usually classified into four groups~\cite{chandra,lavren,kleman}, namely $\lambda^{\pm m}$, $\tau^{\pm m}$, $\chi^{\pm m}$ defect lines ($m$ is the topological charge, either an integer or a half-integer number, positive or negative) and twist disclinations. While disclination lines of type $\chi^{\pm m}$ requires a full 3D system and are not observed here, $\lambda^{\pm m}$ and $\tau^{\pm m}$ defects do occur.

\begin{figure*}[htbp]
\centerline{\includegraphics[width=0.5\textwidth]{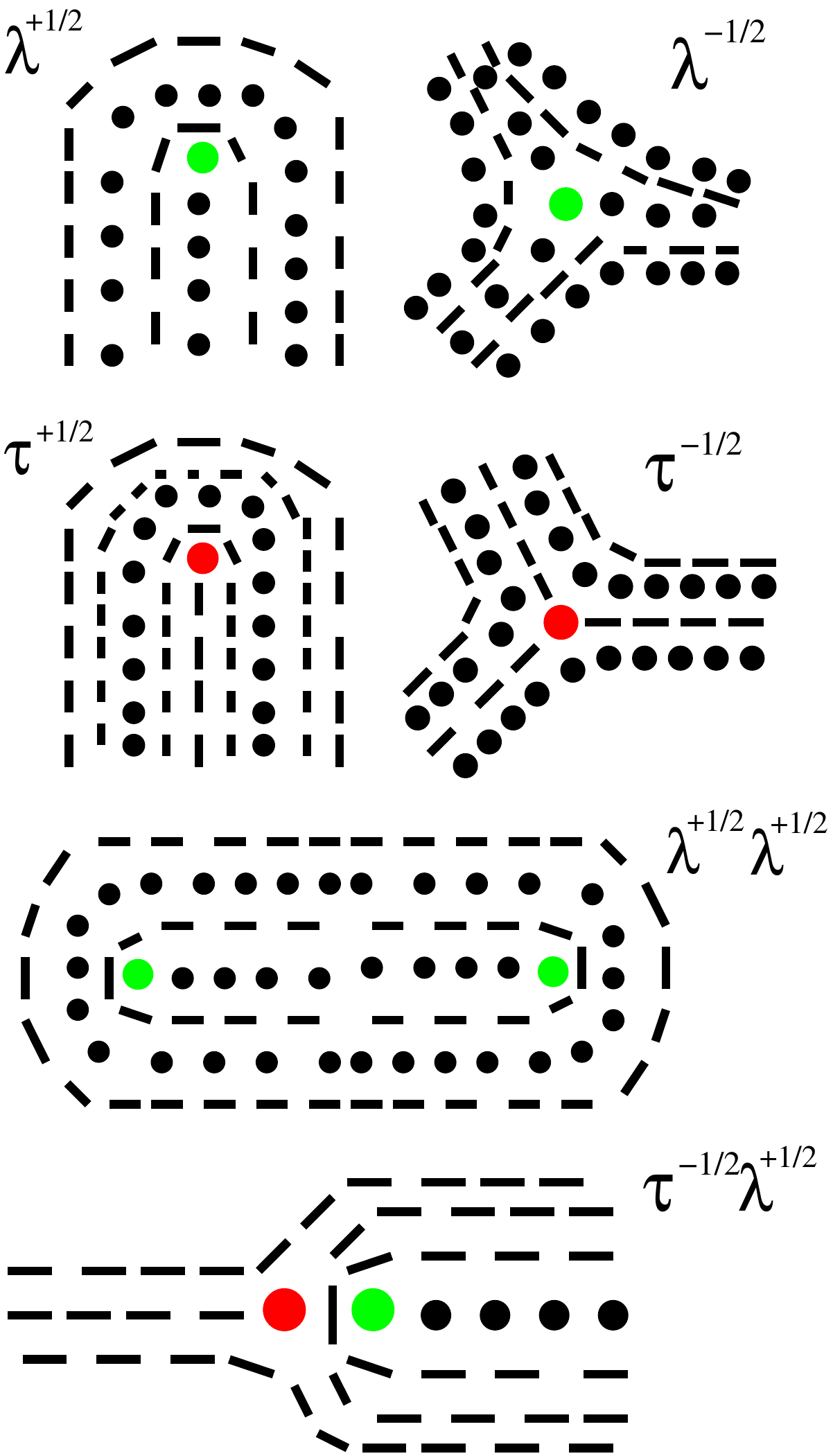}}
\caption{Schematic representation of topological defects $\lambda^{\pm m}$ and $\tau^{\pm m}$ and their combination found in most of our simulations. Black dots and lines represent the local director field, perpendicular to and on the plane, respectively. Unlike green dots, the red ones correspond to a real singularity of the director field~\cite{lavren4,zumer}. An example of $\lambda^{+1}$ defect is that in Fig.~\ref{fig1}b.}\label{fig_def}
\end{figure*}
In a $\lambda$ defect (or disclination line) the director field is well-defined and non-singular: this involves an escape of the director field into the third dimension (for an example, see the centre of the droplet in Fig.~\ref{fig1}b and the first row of Fig.~\ref{fig_def}). Instead, for a $\tau$ defect (or for some of the points in a $\chi$ disclination line) the director field is singular, so that the (in-plane nematic) orientational order drops at the core (see Fig.~\ref{fig_def} second row). As a consequence, we can easily track the location of $\tau$ defects numerically (as we can compute the orientational order in each point), and these are circled in red in our Figures. On the contrary $\lambda$ defects, whose core has radius comparable to the helix pitch $p$, are more difficult to pinpoint exactly, especially in cases where, for reasons of computational efficiency, only a limited number of lattice points can be used to resolve the correct extension of $p$. For this reason, in the text we often refer to a region containing either a $\lambda^{+1}$ defect (or a pair of $\lambda^{+1/2}$ defects) as a ``$+1$-charge region'' (see Fig.~\ref{fig_def} third row). As will be clear from an analysis of the director field patterns in what follows, these two conformations can often morph into one another as the field is switched on or off.

In the case of strong  tangential anchoring and $N=2$ (Fig.~\ref{fig1}b) the helical structure bends significantly in a circular fashion  forming a quasi-planar configuration with the director field that escapes into the third  dimension only in proximity of the droplet center (in this region there is double twist, i.e., twist along all axes lying in the $yz$ plane). As anticipated, this is actually a $\lambda^{+1}$ disclination -- taking into account its topological charge (equal to $+1$), this is required to satisfy global conservation of this quantity.
If, on the other hand, the number of twists increases to $N=4$, the twist energy is too large to be overcome by the bending one and the configuration with the minimal free energy has four defects with topological charge $-1/2$  (lying fully in the $yz$ plane) and three regions of topological charge $+1$ (the stretched stripes in which the director lies along the $x$ direction, 
see Fig.~\ref{fig1}e~\cite{note1}. Note that this structure closely resembles the bipolar structure observed in ~\cite{zumer}.

In Fig.~\ref{fig1}c we show the equilibrium configuration with homeotropic anchoring and $N=2$. In this case a thin $+1$-charge region forms in the center of the droplet (along the diameter in the north/south direction), and is sustained by two twist disclinations of charge $-1/2$ firmly anchored at the interface (see also Fig.~\ref{fig_def}, fourth row). On both sides of this region, the director displays two symmetric splay-bend deformations in order to accommodate the orientation imposed by the homeotropic surface anchoring. This determines the formation of two further defects, $\tau^{1/2}$ disclinations, located symmetrically along the equator. Note that, as required of any $2D$ pattern with these boundary conditions, the total topological charge is therefore again $+1$.
Finally when $N=4$, two more $+1$-charge regions appear symmetrically located with respect to the central one, and eight disclinations form nearby the interface. Two of these are $\tau^{1/2}$ disclinations (those along the equator) whereas the remaining ones are twist disclinations of charge $-1/2$.  Note that in proximity of the two $\tau^{1/2}$ defects, the strong bend distortions of the director field protrude well inside the  center of the droplet, especially for $N=2$.
As previously mentioned, the position of the defects can however be controlled by tuning the value of the elastic constant $K_{lc}$ relative to the anchoring strength $W$. 
For example, with higher values of the ratio $W/K_{lc}$ the defects near the droplet surface are expected to shift towards its centre to pay less surface energy. We will see that the spatial rearrangement of the  defects can be also achieved by using an external electric field.

\section{Switching dynamics of a cholesteric droplet in a  uniform electric field}\label{switching}

In this section we discuss the switching dynamics of a droplet of  cholesteric liquid crystal hosted inside an isotropic fluid, in the presence of a uniform electric field. More specifically, by starting from the equilibrated droplet configurations described in the previous section we switch on a uniform electric field whose direction is either parallel (i.e. along the $y$ axis) or perpendicular (i.e. along $z$) to the helical axis. 
We monitor the dynamics of the system until it reaches a steady state (ON state). We then switch off the electric field and look at the relaxation dynamics towards a zero field (OFF) state. We will show that in general the OFF state is a new metastable state, generally different from the starting equilibrium configuration. The choice of the electric field magnitude $E = \Delta V L$ is crucial.  If $\Delta V \lesssim 2$, the electric field is too weak to alter in an appreciable way the initial equilibrium state. On the other hand if  $\Delta V \gtrsim 5$, the electric field fully destroys the cholesteric phase inside the droplet and a defect-free nematic droplet,  ordered along the field direction ($\epsilon_a$ is positive), is found. Hence we will present only results obtained for intermediate values of $\Delta V$ in which the electric field is sufficiently strong to rearrange and partially change the defect structure but keeping the droplet in a cholesteric-like phase. We first present the results for a droplet with no  anchoring (defect-free droplet) and then move to the more interesting cases in which homogeneous and homeotropic strong anchoring are set.

\subsection{Free-anchoring}\label{free_anch}
\begin{figure*}[htbp]
\centerline{\includegraphics[width=0.77\textwidth]{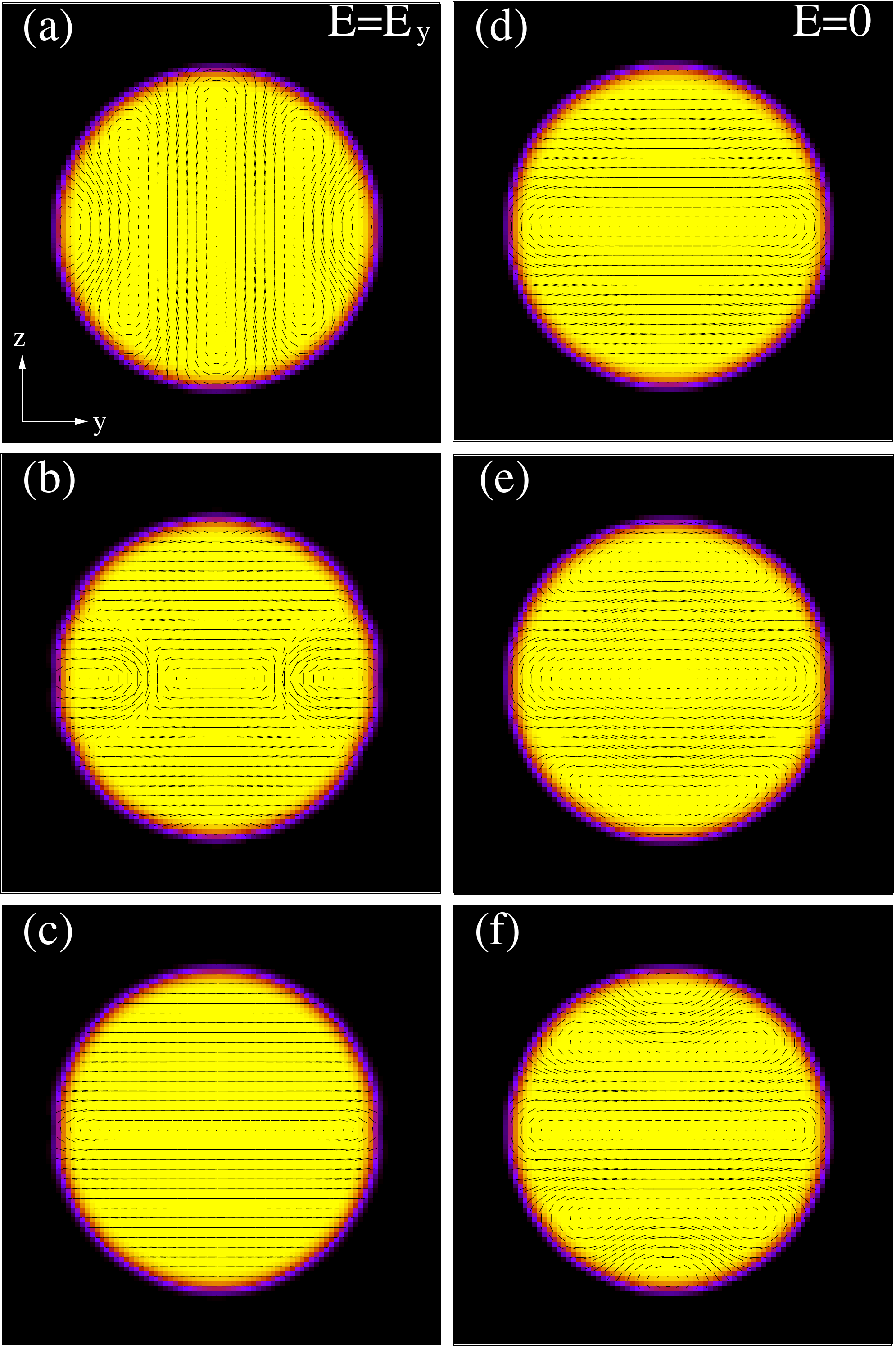}}
\caption{Switching dynamics of a cholesteric droplet with $N=4$ under an electric field applied along the $y$-direction with $\Delta V=5$ and $W=0$ (no anchoring). The field is ON in (a) ($t=4\times 10^5)$,
(b) ($t=4.05\times 10^5$) and (c) ($t=7\times 10^5$) and OFF in (d) ($t=7.01\times 10^5$), (e) ($t=7.05\times 10^5$) and (f) ($t=10^6$). Here $t$ is the simulation time. In particular (c) and (f) represent steady states achieved when the electric field is ON and OFF, respectively. After one switching cycle the director reacquires a cholesteric arrangement with the helix axis parallel to the direction of the electric field.}
\label{fig7A}
\end{figure*}

\begin{figure*}[htbp]
\centerline{\includegraphics[width=1.2\textwidth]{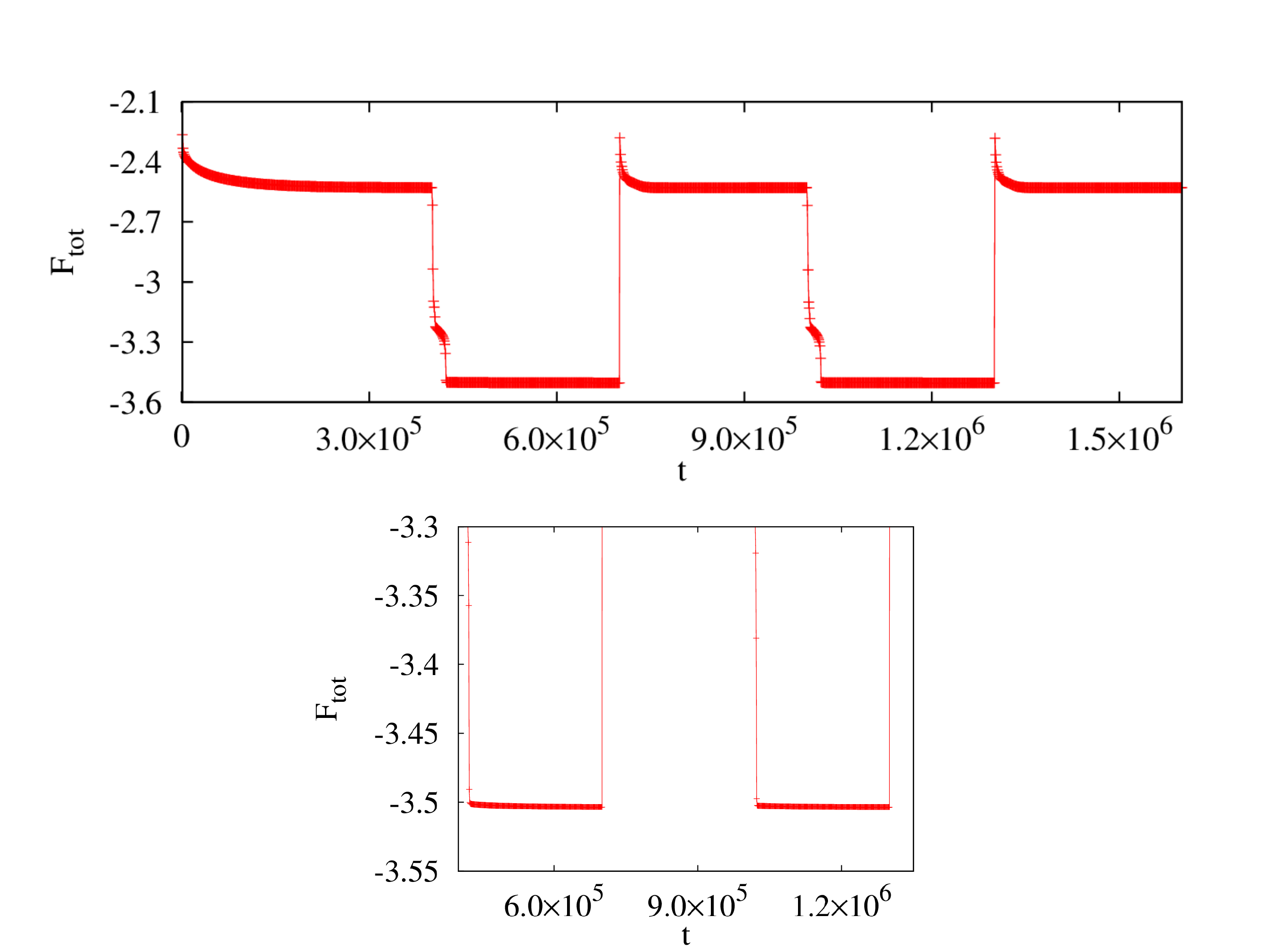}}
\caption{Total free-energy vs simulation time for the $W=0$ case. The field is switched on at $t=4\times 10^{5}$ and switched off at $t=7\times 10^{5}$ during the first cycle. During the second cycle it is switched on at  $t=1\times 10^{6}$ and off at $t=1.3\times 10^{6}$.  Here $t$ is the simulation time. Note that the two minima of the free-energy are equal.}
\label{fig3_}
\end{figure*}

As a reference case we study the switching dynamics for the equilibrium configuration of Fig.~\ref{fig1}d, where a cholesteric phase with  $N =4$ is embedded in a droplet of radius $R$. Since $W=0$ the director field at the droplet surface is free to rotate and no defects (i.e., regions with low orientational order) are present. As one can readily see from Figs.~\ref{fig7A}a-c, as an electric field is switched on along the helix ($y$-axis), the director field gradually aligns along the same direction by following a complex  dynamics involving the formation and annihilation of defect pairs. In steady state (when the field is ON) the droplet is oriented almost completely to yield a 2D  nematic state -- the exception is the equatorial line in which the director field is oriented out of the plane (Fig~\ref{fig7A}c). (A fully nematic order is achieved if a higher value of the electric field is considered). When the electric field is switched off (Fig.~\ref{fig7A}d-f)  the director relaxes to the initial (field-off) equilibrium cholesteric arrangement but with the helix axis now oriented along the $z$-axis. A further ON-OFF cycle, in which the field is first switched on along the $z$-direction, drives the system back to the initial equilibrium state (Fig.~\ref{fig7A}a). The OFF states are of course equivalent as they differ solely by a rotation of the helical axis, and indeed the values of the field-off free-energy configurations are equal (Fig.~\ref{fig3_}).

\subsection{Tangential anchoring}\label{tang_anch}
We now consider the case in which strong tangential anchoring is imposed at the droplet surface. The two equilibrium starting configurations are those reported in Fig~\ref{fig1}b,e with cholesteric helices respectively of $N=2$ and $N=4$ twists. In both cases we have set $\Delta V=3$. The case $N=2$ (Fig.~\ref{fig1}b) is akin to that of a simple nematic droplet. In fact, regardless of the direction of the applied electric field, the director, in the ON steady state, aligns almost everywhere along it, the only exception being a thin stripe passing through the centre of the droplet, in which it orients out of the plane, along the $x$-direction. When the electric field is switched off, the system relaxes back quite rapidly to the initial field-off equilibrium state. 
\begin{figure*}[htbp]
\centerline{\includegraphics[width=0.75\textwidth]{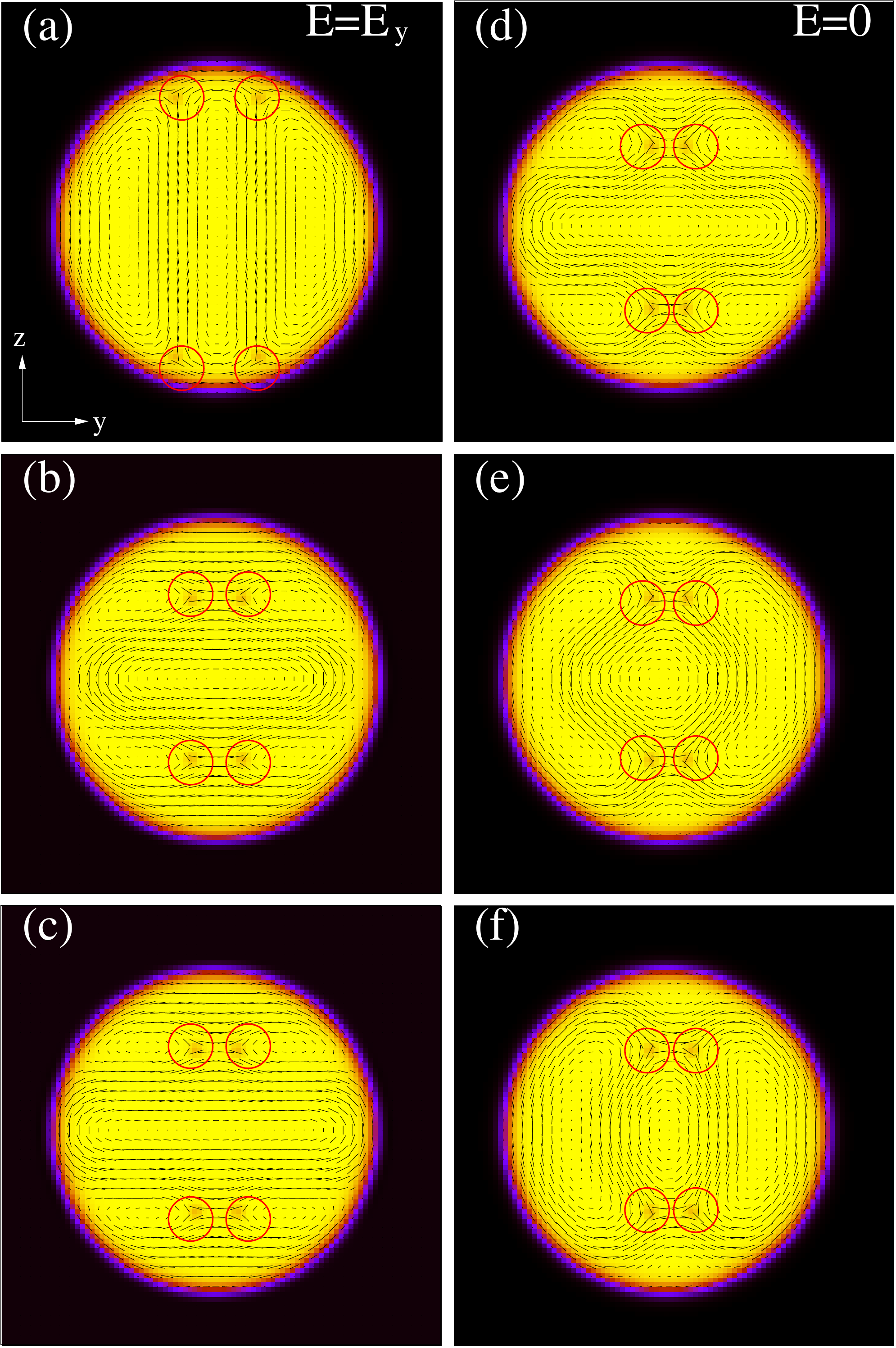}}
\caption{Switching dynamics of a cholesteric droplet with $N=4$ under an electric field applied along the $y$-direction with $\Delta V=3$ and $W=0.04$ (strong tangential anchoring). The field is ON in (a)
($t=4\times 10^5$), (b) ($t=4.5\times 10^5$) and (c) ($t=7\times 10^5$) and OFF in (d) ($t=7.05\times 10^5$), (e) ($t=7.5\times 10^5$) and (f) ($t=10^6$), where $t$ is the simulation time. In particular (c) and (f) represent steady states achieved when the electric field is ON and OFF, respectively. At the end of the cycle the droplet is stuck into a novel metastable state achieved through a complex internal rearrangement of the topological 
defects.}\label{fig2}
\end{figure*}

A richer behaviour is observed for the cholesteric phase with  $N=4$ and when the electric field is applied along the helix ($y$) axis (see Fig.~\ref{fig2}). During the ON dynamics, as the director starts to align along the direction of the electric field,  the four defects abandon the droplet surface and move towards the inner region of the droplet. While the director field near the droplet surface remains almost everywhere tangential, inside the droplet it undergoes pronounced symmetric bend distortions that move along the $y$ axis in opposite directions (Fig.~\ref{fig2}a-b). At the steady state the bend distortions are confined near the interface and in between the defects, while the director is almost fully parallel to the $y$-direction in the bulk, with the exception of the equatorial line and of the opposite sides of the in-plane topological defects, in which it is oriented off plane (see Fig.~\ref{fig2}c). These patterns are associated with three $+1$-charge regions.
Note that in the ON steady state the initial equilibrium  helical order (with axis along the $y$-direction) has been replaced by a novel arrangement in which a cholesteric-like structure, with the axis now along the $z$-direction, emerges. 

\begin{figure*}[htbp]
\centerline{\includegraphics[width=1.1\textwidth]{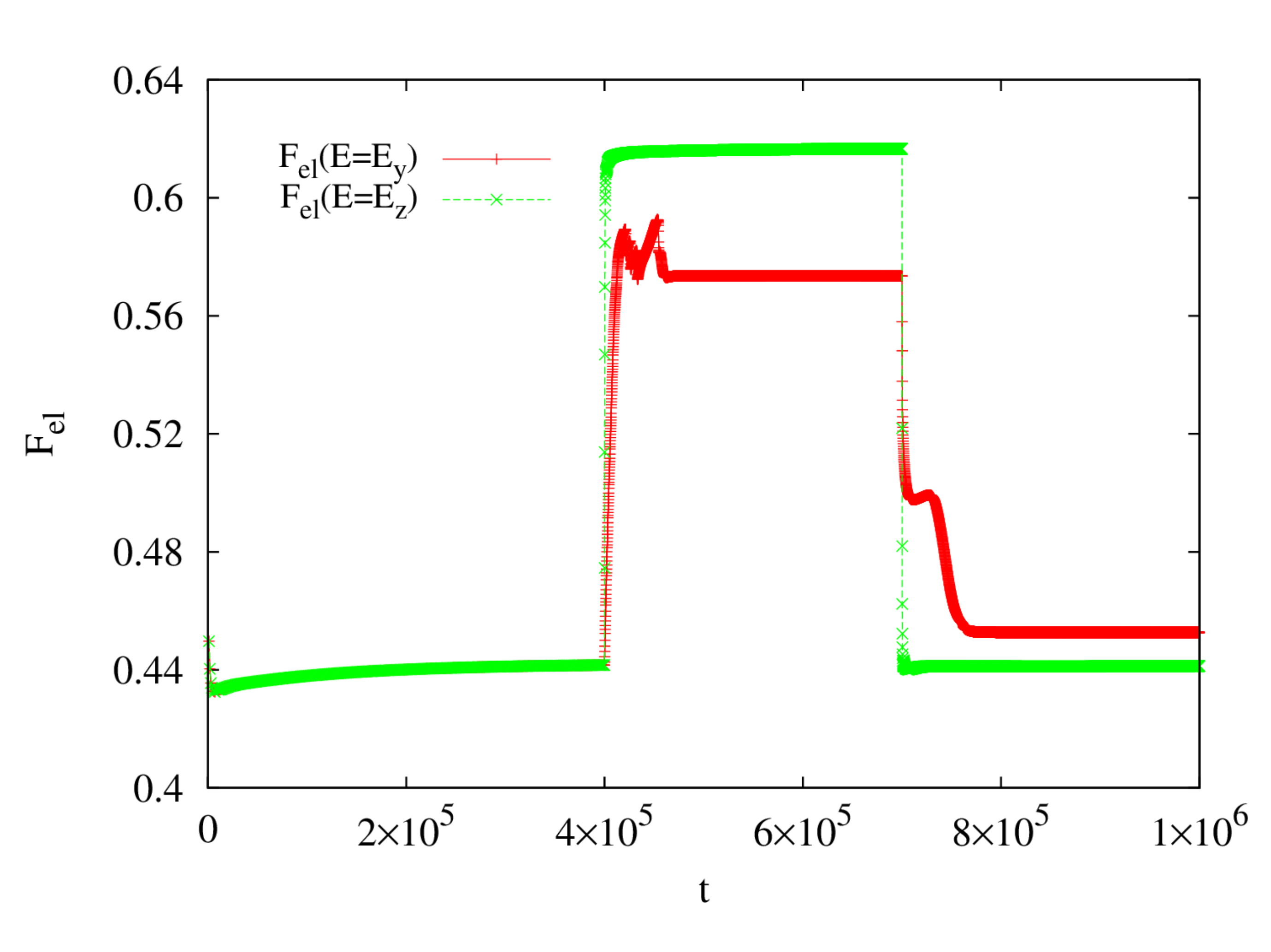}}
\caption{Elastic free energy of the droplet during a switching dynamics, with the electric field applied along  $y$-direction (red curve, plusses) with voltage $\Delta V_{y}=3$ and along the $z$-direction  (green, crosses) with voltage $\Delta V_{z}=5$, both for $N=4$ ($q_{0}=2\pi/32$) and strong tangential anchoring ($W=0.04$). The field is switched on at simulation time $t=4\times 10^5$ and switched off at $t=7\times 10^5$. In the former case the final field-off state has higher elastic free-energy than the initial one.}
\label{fig3}
\end{figure*}

When the field is switched off, the director gradually relaxes towards a new steady state in which the helical arrangement is almost restored. While the position of the in-plane defects remains overall unaltered with respect to that achieved at the ON state, the position of the three $+1$-charge regions is restored almost in the configuration they had in the equilibrium field-off state, as they follow the rearrangement of the director field. This new OFF state is metastable, as it is characterised by an elastic free energy (i.e. the term proportional to  $K_{lc}$ in Eq.(\ref{free_E})) higher than that of its equilibrium counterpart (see Fig.~\ref{fig3}, red/plus symbols). 
The higher value of the free energy achieved by this OFF steady state is due to the additional elastic energy necessary to bend the director in proximity of the defects.

If the electric field is switched along the $z$-direction, the dynamics observed is overall less rich than the previous case. Here we set $\Delta V=5$, as for lower values of the electric field the effects on the defects dynamics are really  negligible. After the electric field is switched on, the in-plane defects are soon pushed towards the interface of the droplet as the director aligns parallel to the $z$-direction, while the $+1$-charge regions slightly stretch along the field direction. When the field is switched off the in-plane defects migrate back to the initial position driving the system to a new final metastable state (see the elastic free energy in Fig.~\ref{fig3}, green/crosses symbols), whose director arrangement is similar to the initial equilibrium state.

\subsection{Homeotropic anchoring}\label{perp_anch}

\begin{figure*}[htbp]
\centerline{\includegraphics[width=0.74\textwidth]{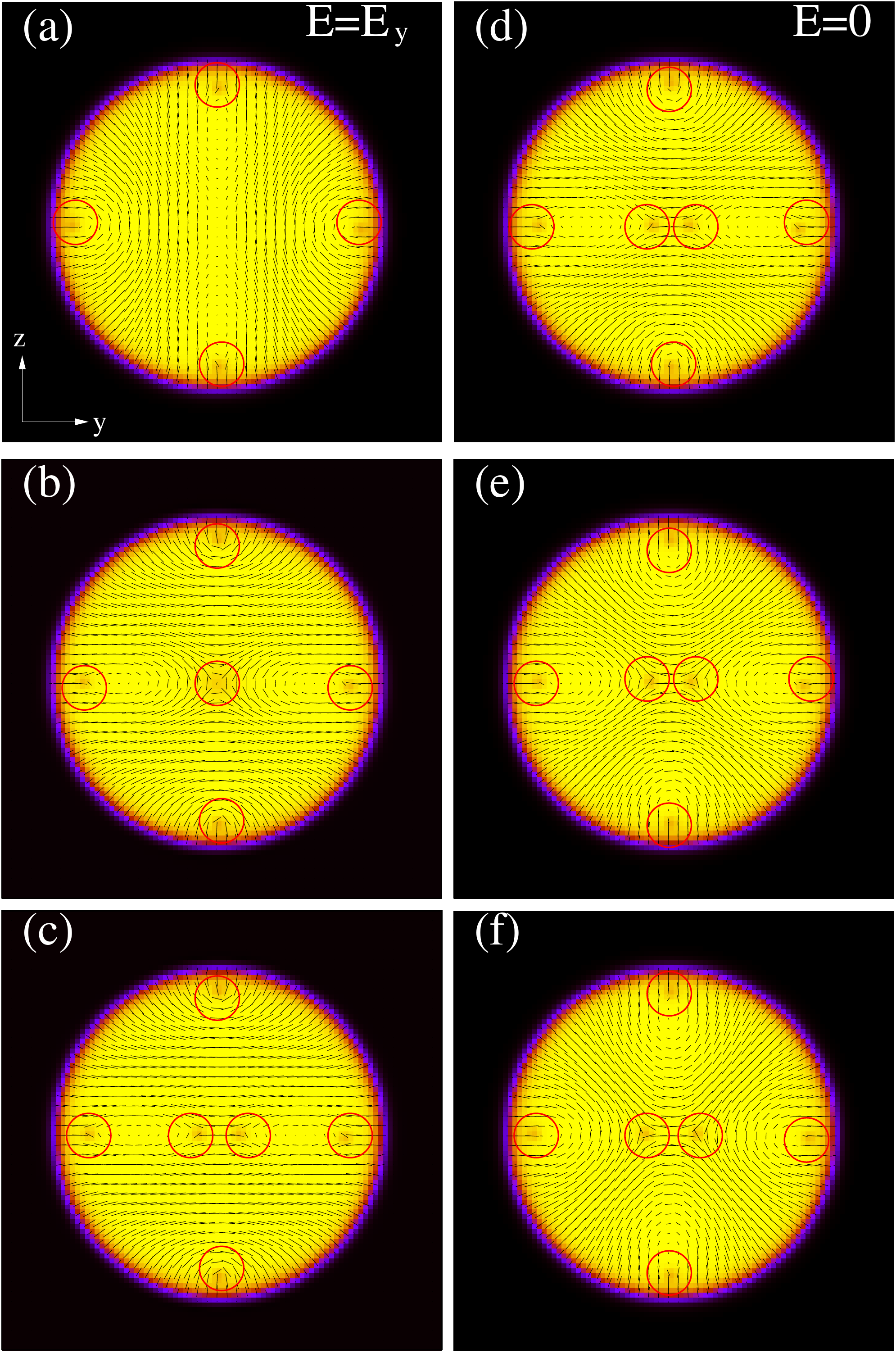}}
\caption{Switching dynamics of a cholesteric droplet with $N=2$ under an electric field applied along the $y$-direction with $\Delta V=2.5$ and $W=-0.04$ (homeotropic anchoring). The field is ON in (a)
($t=4\times 10^5$), (b) ($t=4.15\times 10^5$) and (c) ($t=7\times 10^5$) and OFF in (d) ($t=7.02\times 10^5$), (e) ($t=7.08\times 10^5$) and (f) ($t=10^6$), where $t$ is the simulation time. In particular (c) and (f) represent steady states achieved when the electric field is ON and OFF, respectively. After one switching cycle two in-plane $-1/2$ defects form in the centre of the droplet sustained by intense bend distortions induced by the surface homeotropic anchoring.}
\label{fig7}
\end{figure*}

A much richer phenomenology is observed when the surface anchoring is homeotropic. Let us consider first the case with $N=2$ where the defect arrangement of the equilibrium configuration (Fig~\ref{fig1}c) 
is simpler. Its switching on-off dynamics is reported in Fig.~\ref{fig7} when the electric field is applied along the $y$-direction. One can notice that during the first half cycle in which the electric field
 is on (Fig.~\ref{fig7}a-c), while the director aligns along the direction of ${\bf E}$, a temporary in-plane defect of charge $-1$ emerges at the  centre of the droplet. This defect is unstable and later on 
it  splits into two twist disclinations of charge $-1/2$ each. 
The initial $+1$-charge region switches to two separate double twist regions, separated by a hyperbolic hedgehog in the centre (whose topological charge is $-1$).
Also note that, due to the alignment of the director field  along the direction of  ${\bf E}$ the defects near the surface change their topological charge and their nature. The two $\tau^{1/2}$ defects along 
the equator became two twist disclinations of charge $-1/2$, whereas the two twist disclinations of charge $-1/2$ (located along the north/south direction) turn into two  $\tau^{1/2}$ defects.
Although the switching-off protocol has negligible effects on the defect dynamics, it favours the formation of two large symmetric splay-bend deformations spanning the whole droplet (Fig.~\ref{fig7}d-f). The 
defects nearby the interface return to their initial field-off configuration whereas those in the bulk turn into two fully in-plane $-1/2$ defects. Note that the final field-off state (f) is always
restored if a further switching on-off cycle, starting from (f) itself, is applied.

\begin{figure*}[htbp]
\centerline{\includegraphics[width=0.73\textwidth]{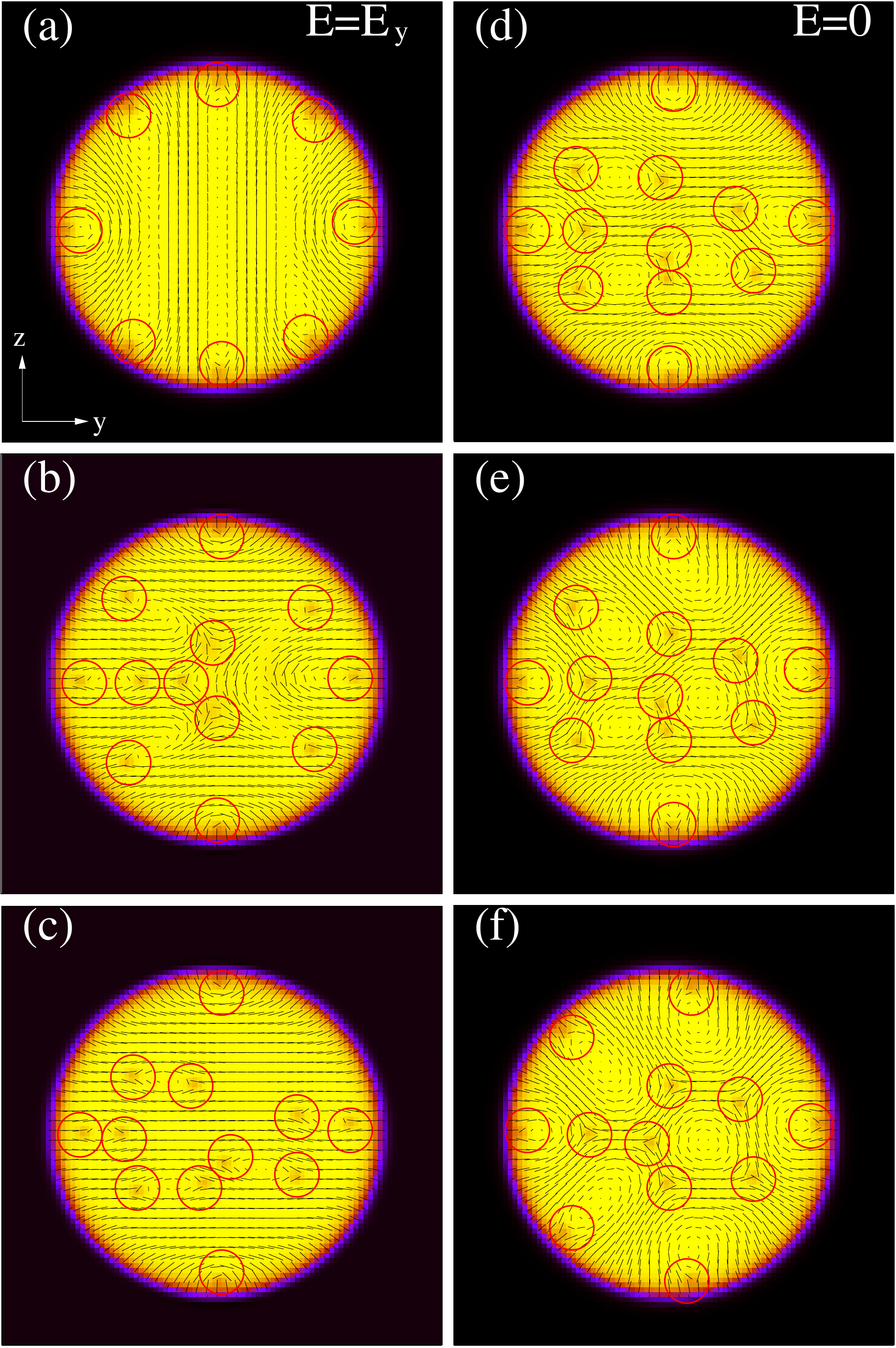}}
\caption{Switching dynamics of a cholesteric droplet with $N=4$ under an electric field applied along the $y$-direction with $\Delta V=5$ and $W=-0.04$ (strong homeotropic anchoring). The field is ON in (a) ($t=4\times 10^5$), (b) ($t=4.03\times 10^5$) and (c) ($t=7\times 10^5$) and OFF in (d) ($t=7.01\times 10^5$), (e) ($t=7.08\times 10^5$) and (f) ($t=10^6$), where $t$ is the simulation time. In particular (c) and (f) represent steady states achieved when the electric field is ON and OFF, respectively. A complex dynamics is observed during the entire switching cycle. The initial symmetric defect arrangement is destroyed by the electric field and, during the off dynamics, the droplet relaxes towards a new metastable state with several defects.}
\label{fig4}
\end{figure*}

The case with $N=4$ starts from an equilibrium configuration characterised by a large number of defects, which makes the switching dynamics more complex due to the onset of new metastable states. This is expecially true when the electric field ($\Delta V=5$) is applied along the helix axis (see Fig.~\ref{fig4}).
Indeed after the field is switched on (Fig.~\ref{fig4}a-c), the director undergoes a complex reorientation dynamics ending with an almost fully nematic ON state (along the $y$-direction) decorated  with several defects. In particular several pairs of twist disclinations of charge $-1/2$ are anchored to $+1$-charge regions (and usually are pinned there by $\lambda^{+1/2}$ defects immediately nearby each of the defect). These structures span the bulk of the droplet, while two $\tau^{1/2}$ disclinations are stuck at the interface. Note that here higher values of the reduced potential $\Delta V$ would have driven the director toward a fully aligned nematic state. After the field is switched off (Fig.~\ref{fig4}d-f), the liquid crystal relaxes towards a new metastable state in which some defects cluster deep inside the droplet while the remaining ones are located close to the interface. The director, in turn, develops a more intricate profile, in which pronounced bend distortions in the  vicinity of the defects are spaced out by $+1$-charge regions. Note that, if starting from this state the same switching cycle is applied once again, 
the system gets stuck into a field-off state that is slightly different from the one shown in Fig.~\ref{fig4}f.
The switching dynamics therefore strongly influences the kinetic pathway followed by the system 
and, at least in this case, impedes an easy restoration of the initial equilibrated states, such as 
the symmetric structure of the director field  of Fig.~\ref{fig4}a, which is, hence, not 
easily reproducible. This state is also unstable to weak perturbation: an alternative switching dynamical pathway can be observed if a weak random perturbation   is initially added, for instance, to the order parameter ${\bf Q}$ (about $10\%$ of its equilibrium value).

\begin{figure*}[htbp]
\centerline{\includegraphics[width=0.77\textwidth]{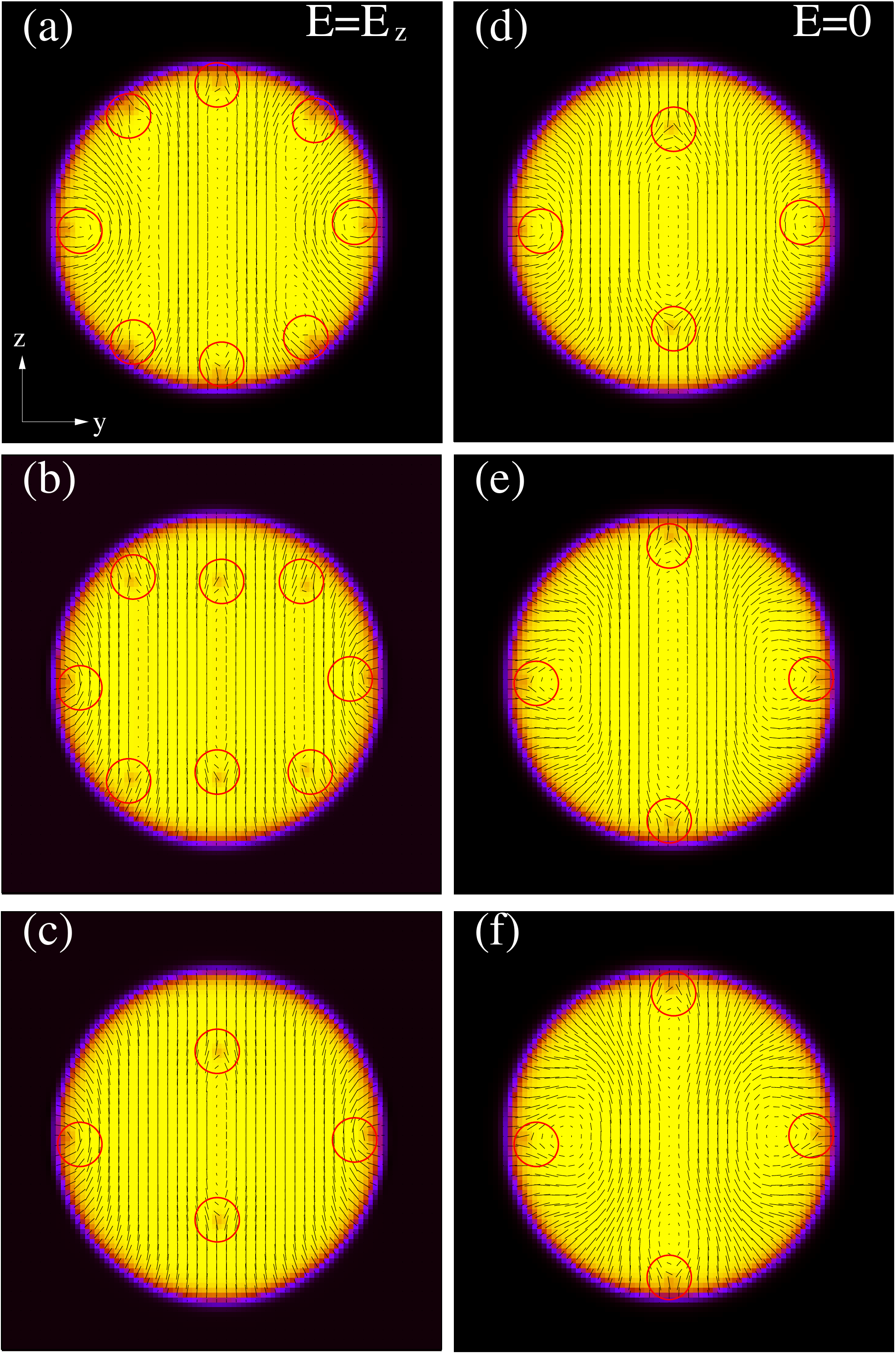}}
\caption{Switching dynamics of a cholesteric droplet with $N=4$ under an electric field applied along the $z$-direction with $\Delta V=5$ and $W=-0.04$ (homeotropic anchoring). The field is ON in (a) ($t=4\times 10^5$), (b) ($t=4.1\times 10^5$) and (c) ($t=7\times 10^5$) and OFF in (d) ($t=7.01\times 10^5$), (e) ($t=7.08\times 10^5$) and (f) ($t=10^6$), where $t$ is the simulation time. The defect dynamics is overall simpler than the previous case (see Fig.~\ref{fig4}); indeed a cholesteric-like fashion of the director field is obtained after one switching cycle.}
\label{fig5}
\end{figure*}

\begin{figure*}[htbp]
\centerline{\includegraphics[width=1.1\textwidth]{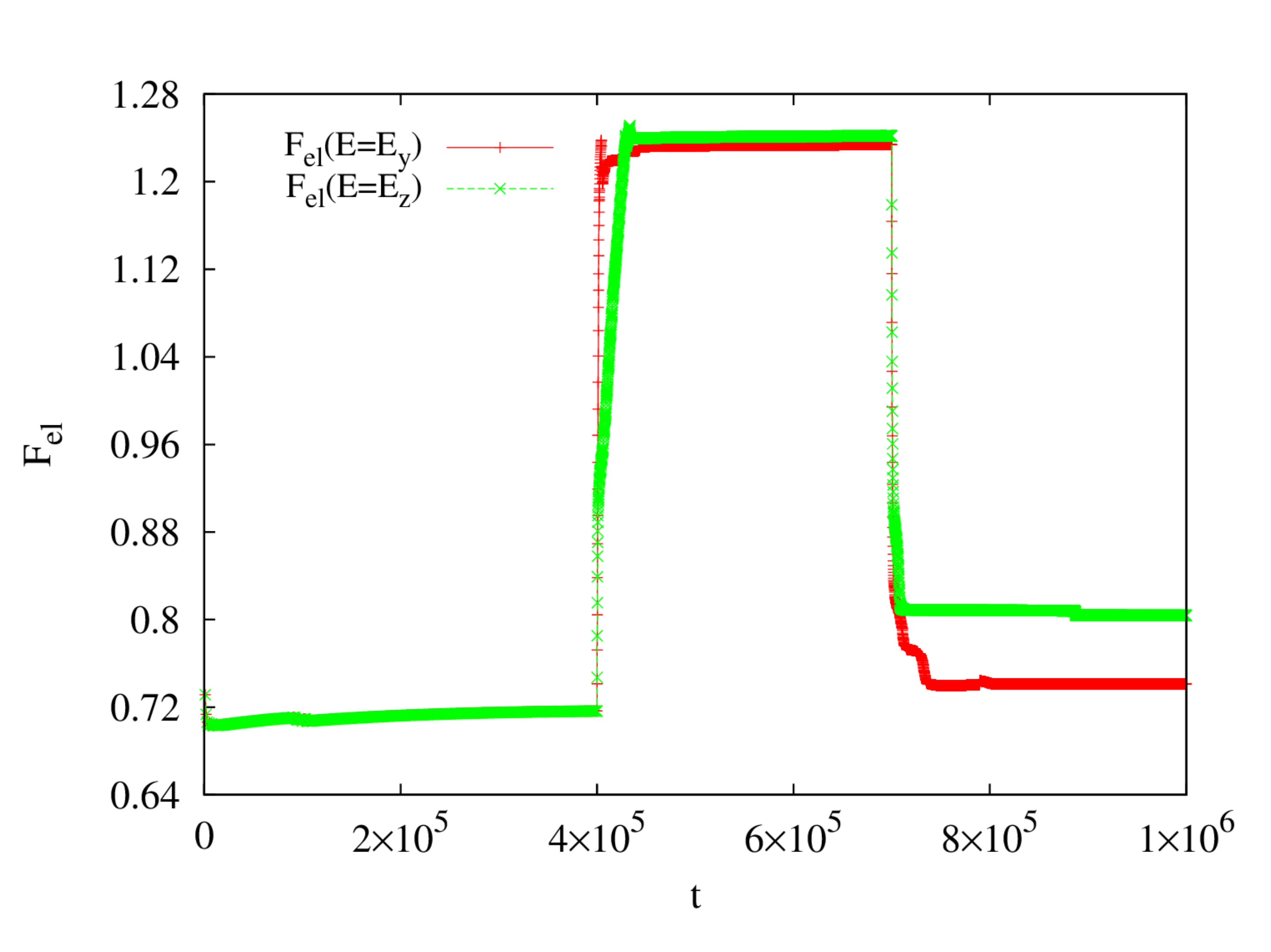}}
\caption{Elastic free energy of the system when a uniform electric field is switched along $y$-direction (red curve) with voltage $\Delta V_{y}=5$ and along the $z$-direction (green) with voltage $\Delta V_{z}=5$, both for $N=4$ ($q_{0}=2\pi/32$) and strong homeotropic anchoring ($W=-0.04$). The field is switched on at simulation time $t=4\times 10^5$ and switched off at $t=7\times 10^5$. The field-off states (before and after one switching cycle) have different values in both cases.}
\label{fig6}
\end{figure*}

If the field is applied along the $z$-direction, a less complex, but more intriguing dynamics is observed. During the ON dynamics, while the director progressively aligns along the direction of the electric field, the defects located symmetrically in the north and south part of the droplet move towards its centre, leaving a characteristic straight stripe signature which identifies regions where the director lies along the $x$-direction (Fig.~\ref{fig5}a-c). 
If the electric field is strong enough the final ON state is an almost nematic phase aligned along the $z$-direction, in which a central stripe with out-of-plane director orientation is sustained by two twist disclinations of charge $-1/2$. The two $\tau^{1/2}$ defects that survive in the ON state are those located along the equatorial line, whose position was unaffected by the switching on dynamics (Fig.~\ref{fig5}c). During the  switching OFF relaxation  (Fig.~\ref{fig5}d-f), the two defects inside the droplet move back to the interface while the director gradually relaxes into a final state in which splay-bend distortions emerge as a result of the strong homeotropic anchoring. Notice that, regardless of the direction of the applied field, after one cycle the liquid crystal gets stuck into a metastable state, whose elastic free-energy is higher than the equilibrium one (Fig.~\ref{fig6}) and is very similar to the equilibrium one for $N=2$ (Fig.~\ref{fig1}e).

One might ask whether, by starting from one of these states, a bistable behaviour may be observed. Bistability would require finding a way back to the equilibrium state (Fig.~\ref{fig5}a) starting from the metastable state in Fig.~\ref{fig5}f. However, our simulations suggest that this is not the case, as, typically, the field-off state (Fig.~\ref{fig5}f) is either restored after a further on-off cycle if the switching on is along $z$, or rotates by 90 degrees if the switching on is along $y$. It is possible that decreasing the elastic constant would lower the energy barrier, hence making it more easily to bypass it, with a high enough voltage.

\subsection{Switching dynamics for a cholesteric phase within planar walls}\label{planar_walls}

The results discussed so far have addressed the physics of a cholesteric liquid crystal when confined into a circular domain, such as that of a droplet. One may wonder how much of the phenomenology just described is due to the geometry of the confined region. In particular it is interesting to pinpoint the role played by the curvature of the boundary in the switching dynamics by looking at the simpler case of a cholesteric phase confined between straight walls. This issue is also of interest for practical applications, such as in flexible displays in which an improvement of the viewing angle (namely the limited range of angles from which the display is readable) is typically achieved by a proper control of the defect dynamics~\cite{apl}.

\begin{figure*}[htbp]
\centerline{\includegraphics[width=0.86\textwidth]{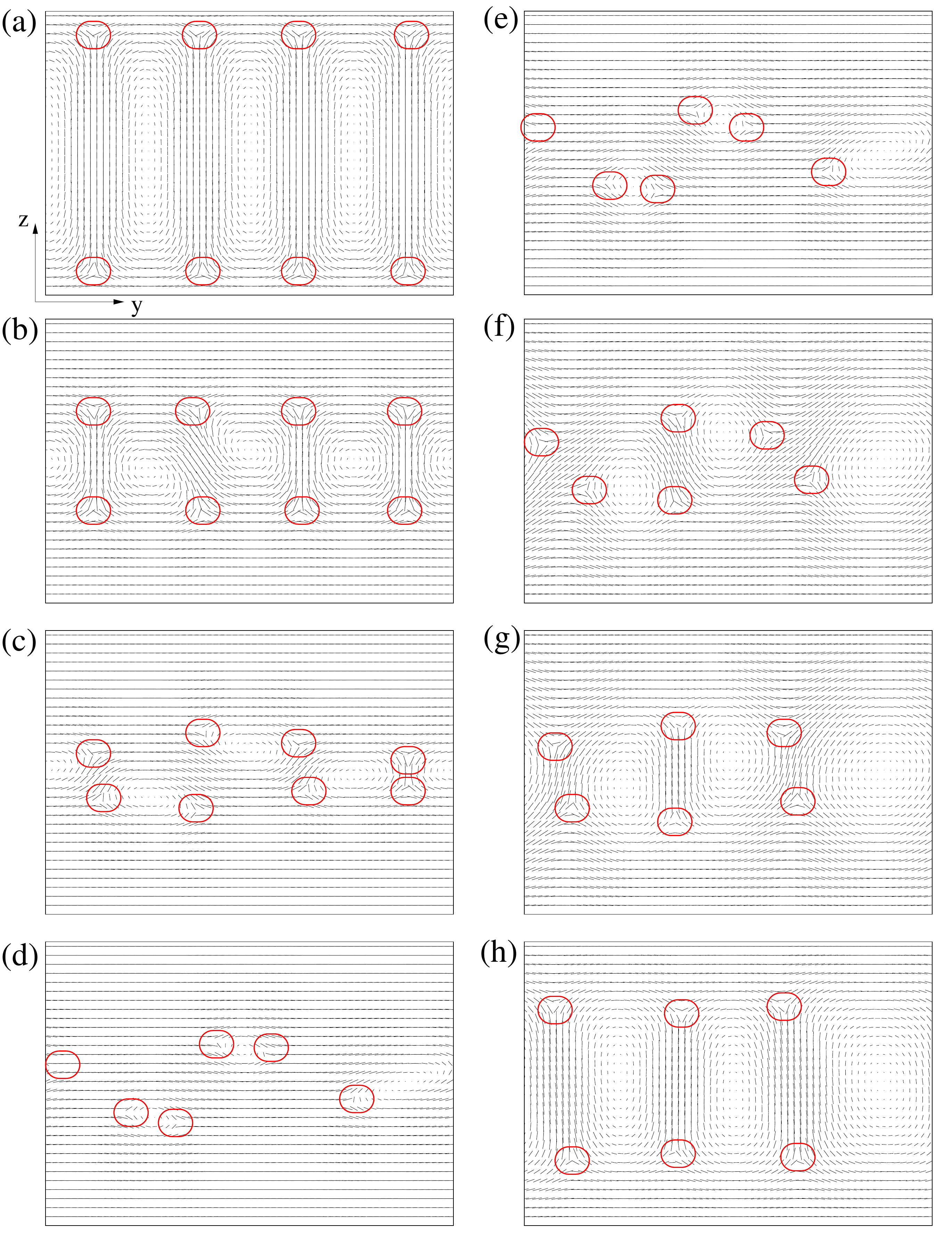}}
\caption{Switching dynamics of a cholesteric liquid crystal confined between two flat solid walls and under an electric field applied along the $y$-direction, with $\Delta V=3$, $\alpha_s=0.4$ (strong tangential anchoring), $L_z=64$ and $L_y=128$. The field is ON in (a) ($t=10^5$), (b) ($t=1.05\times 10^5$), (c) ($t=1.1\times 10^5$) and (d) ($t=4\times 10^5$) and OFF in (e) ($t=4.01\times 10^5$), (f) ($t=4.05\times 10^5$), (g) ($t=4.1\times 10^5$) and (h) ($t=7\times 10^5$), where $t$ is the simulation time. In particular (d) and (h) represent steady states achieved when the electric field is ON and OFF, respectively. After one switching cycle a cholesteric arrangment is almost restored in the bulk of the cell, but not close to the boundaries.}
\label{fig8}
\end{figure*}

\begin{figure*}[htbp]
\centerline{\includegraphics[width=1.1\textwidth]{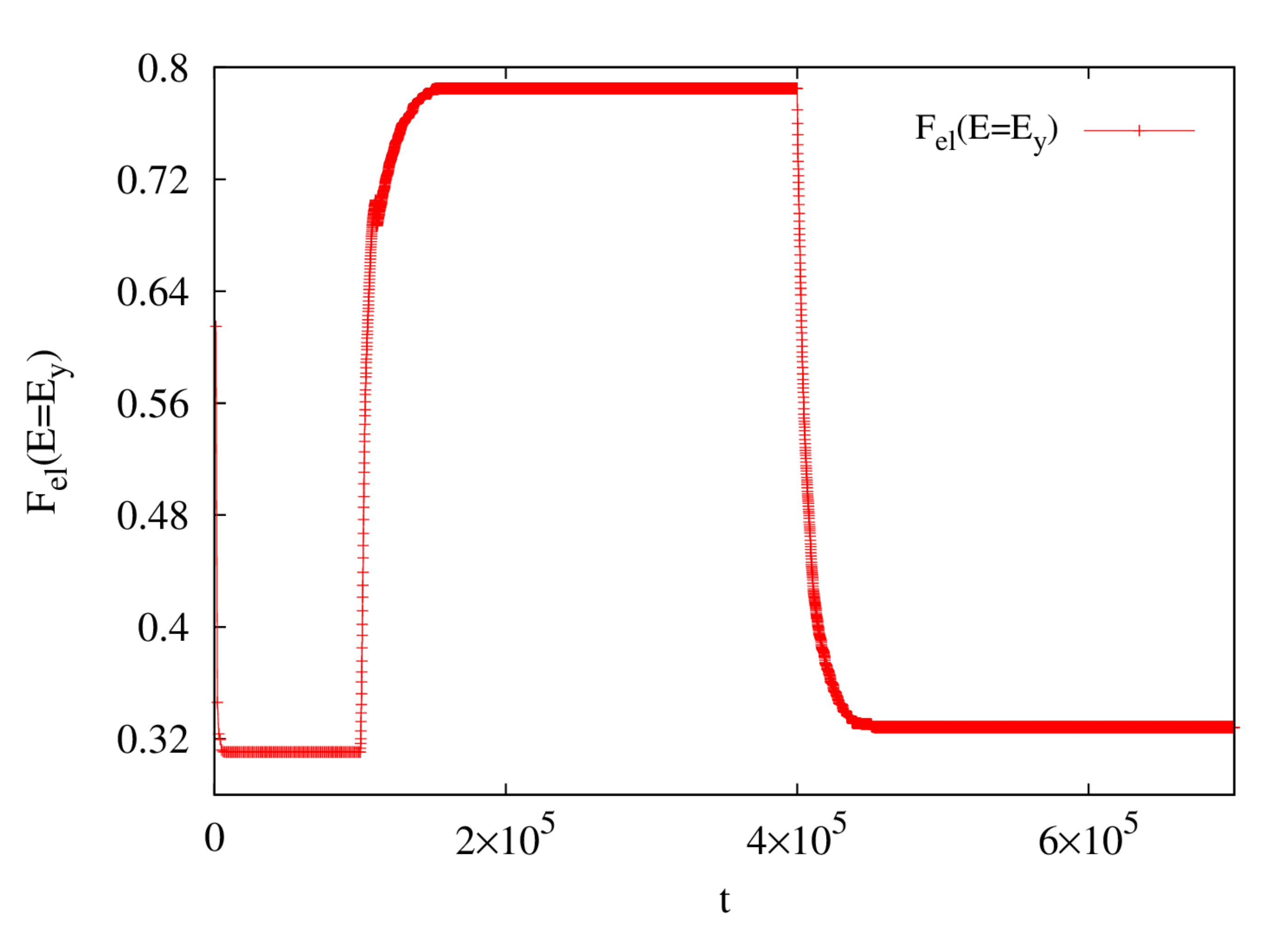}}
\caption{Elastic free energy of cholesteric liquid crystal when confined between flat walls. The electric field is switched on along the $y$-direction (with $\Delta V_{y}=3$) at simulation time $t=10^5$, and switched off at $t=4\times 10^5$.}
\label{fig10}
\end{figure*}

To clarify this point  we have simulated the field-induced switching dynamics for a cholesteric phase sandwiched  between two parallel walls located at $z=0$ and $z=64$ lattice units. Along the $y$-direction, periodic boundary conditions are considered. The equilibrium configuration  corresponds to a full cholesteric phase (namely $\phi\simeq \phi_0$ everywhere) with helix axis along the $y$-direction and $N=4$ (see Fig.~\ref{fig8}a). Strong tangential anchoring has been set at the walls as well as no-slip condition for the fluid velocity (${\bf u}_w=0$). Because of the conflict orientation between the cholesteric phase at the boundary and the tangential anchoring, this set up favours the formation of eight defects of topological charge $-1/2$ pinned at the walls and four $+1$-charge regions in the bulk (each of these has two $\lambda^{+1/2}$ defects at its boundary). Note that, comparing this structure with the droplet of Fig.~\ref{fig1}a, the planar wall set up can accomodate all the eight defects expected for a helix with $N=4$ twist. When the field is switched on along the $y$-direction, defects located on opposite sides approach each other, as the director aligns parallel to ${\bf E}$ by starting from the wall and then penetrating into the bulk (Fig.~\ref{fig8}b). The defect motion  persists until they get stuck roughly at the middle  of the cell (Fig.~\ref{fig8}c-d), by which time each has rotated by almost $90$ degrees. 
When the field is switched off (Fig.~\ref{fig8}e-h), the defects rotate and migrate back towards the wall; however they get stuck before reaching it, so that the cholesteric only partially recovers its twisted arrangement with an elastic free energy different from the initial one (see Fig.~\ref{fig10}). 
This is because there is not enough driving force to twist the director field close to the walls, where there is strong tangential anchoring. 
Therefore, while within the droplet a cholesteric order is restored almost completely after a switching cycle (see Fig.~\ref{fig2}f), with flat walls this is not the case: in this case, curvature facilitates going back to the equilibrium (field off) pattern. Indeed, the presence of regions of different orientation of the director field (due to the presence of topological defects) appears to be more favoured within higher curvature domains. Potentially, this may be important for device design, as it is often desirable to have multi-domain devices (e.g., to enhance the viewing angle): our results suggest that these may be easier to achieve for a liquid crystal droplet than for a standard sample sandwiched by solid walls.
Finally, it is worth investigating the switching dynamics when the cholesteric axis is perpendicular to the walls.
In this case the equilibrated director profile at zero field is a defect-free cholesteric liquid crystal. 
Its arrangement is fully retained if the anchoring is tangential while it is preserved only within the cell if the anchoring is perpendicular. In the latter case one observes pronounced bend distortions near the walls 
favouring a nematic-like order in their proximity. The switching ON dynamics generally follows two possible pathways. For sufficiently high voltage a distortion-free nematic order is found, in which the director is 
oriented parallel to the applied field. For intermediate values of the voltage the cholesteric 
structure is generally preserved. The switching OFF dynamics instead generally restores a cholesteric order.
Note that the dynamics of this set up is overall less complicated than the previous one due to the 
absence of topological defects which usually alter both velocity and director orientation fields.

\section{Dynamics under a rotating electric field}\label{rotate}

In the previous section we have described the switching dynamics of a cholesteric droplet when a uniform electric field is applied along a predefined direction. It is also of interest the study of the dynamics in the presence of a time-dependent non-uniform electric field, such as a rotating one at constant frequency $\omega$.
Experimental realizations of rotating cholesteric droplets range from those whose motion is triggered by the gradient of an external field, such as temperature~\cite{Lehmann,Oswald,Oswald2,Oswald22} or electric potential~\cite{Pratibha,Pratibha2}, to molecular rotors, such as proteins driving self-organization in the cytoskeleton~\cite{Sykes}. 
In this study we adopt a simple implementation where a cholesteric droplet is subject to an electric field periodically rotating at constant frequency $\omega$. This is achieved by setting  $E_{y}=-\Delta V_ysin(\omega t)/L_y$ and $E_{z}=\Delta V_zcos(\omega t)/L_z$, with $\omega>0$, which determines a counterclockwise rotation. 
Like the previous cases, the reduced potential ranges between $2\lesssim \Delta V\lesssim 5$, as for lower values the director remains almost unaltered whereas for higher values the usual ordered nematic state is achieved. 
Note that this study is different from those in which the rotation results from the presence of an electric field (often described as the electric Lehmann effect~\cite{Lehmann,Padmini}), as this arises due to an explicit torque term, proportional to ${\bf n}\times {\bf E}$,  in the expression of the  molecular field of the LE theory~\cite{Oswald3}, which we do not include here.

\begin{figure*}[htbp]
\centerline{\includegraphics[width=0.57\textwidth]{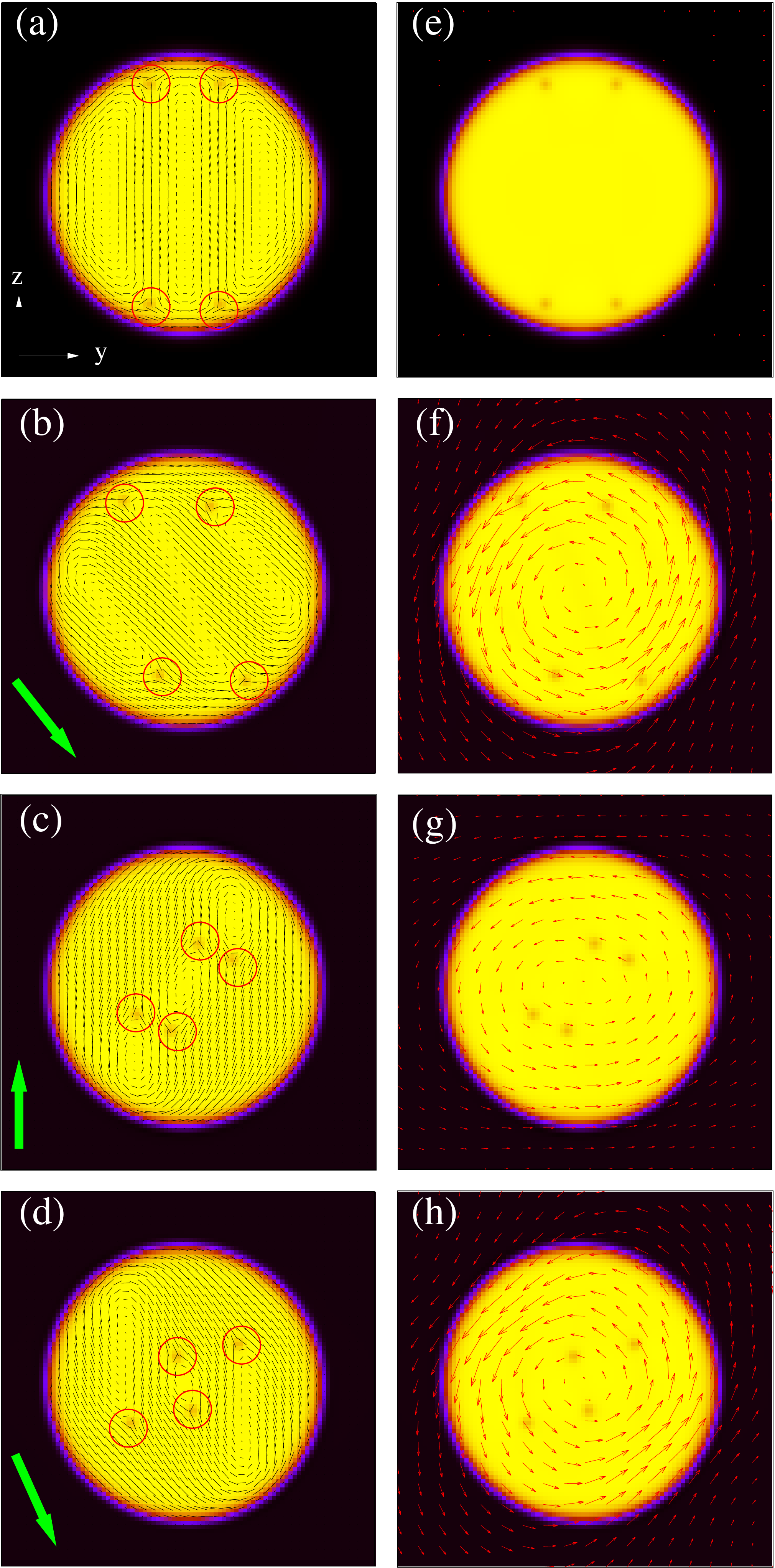}}
\caption{Dynamics of a cholesteric droplet (with $N=4$) under a counterclockwise rotating electric field. The reduced potential is with $\Delta V_y= \Delta V_z=4$, $\omega=5\times 10^{-3}$ and $W=0.04$ (strong tangential anchoring). The left column displays director and defect dynamics; the right column displays the velocity field. The green arrow indicates the direction of the electric field. Corresponding simulation times are $t=4\times 10^5$ for (a)-(e), $t=4.1\times 10^5$ for (b)-(f), $t=4.3\times 10^5$ for (c)-(g), and  $t=4.75\times 10^5$ for (d)-(h)}
\label{fig11}
\end{figure*}

In Fig.~\ref{fig11} we show the dynamics of a cholesteric droplet subject to a rotating electric field with  $\Delta V_y= \Delta V_z=4$, $\omega=5\times 10^{-3}$ and with strong tangential anchoring at the droplet surface (see also movie S1 for the full dynamics). The initial configuration  is the equilibrium one reported in Fig.~\ref{fig1}e for a cholesteric phase with $N=4$ and helix axis along the $y$-direction. 
Under the action of the electric field the four defects, initially located at the droplet surface (Fig.~\ref{fig11}a) start to  rotate counterclockwise (Fig.~\ref{fig11}b) and then move towards the centre of the droplet. The $+1$ charge regions also reshape, and the cholesteric layers temporarily align along a common direction (Fig.~\ref{fig11}c) and then merge.
The electric field, in turn, tends to destroy the initial cholesteric order, and to favour the formation of regions in which the director is aligned along its direction (Fig.~\ref{fig11}c-d), except nearby the interface where strong anchoring keeps the director fixed. 
Notice that, despite the periodicity of the electric field direction ($\omega$ is constant), the defect motion is not periodic, although characteristic patterns frequently appear (see Fig.~\ref{fig11}b,c,d). 
The shape of the $+1$-charge regions (either circular or stripe-like) depends, in turn, on the mutual position of the defects, typically located either in the bulk or near the interface.
The corresponding velocity field is reported in Fig.~\ref{fig11}e-h. This is characterized by a vortex pattern that is almost negligible at the droplet centre  and higher in proximity of the interface, where $|{\bf u}|_{max}\simeq 4.7\times 10^{-4}$ (calculated in simulation time). 

\begin{figure*}[htbp]
\centerline{\includegraphics[width=0.65\textwidth]{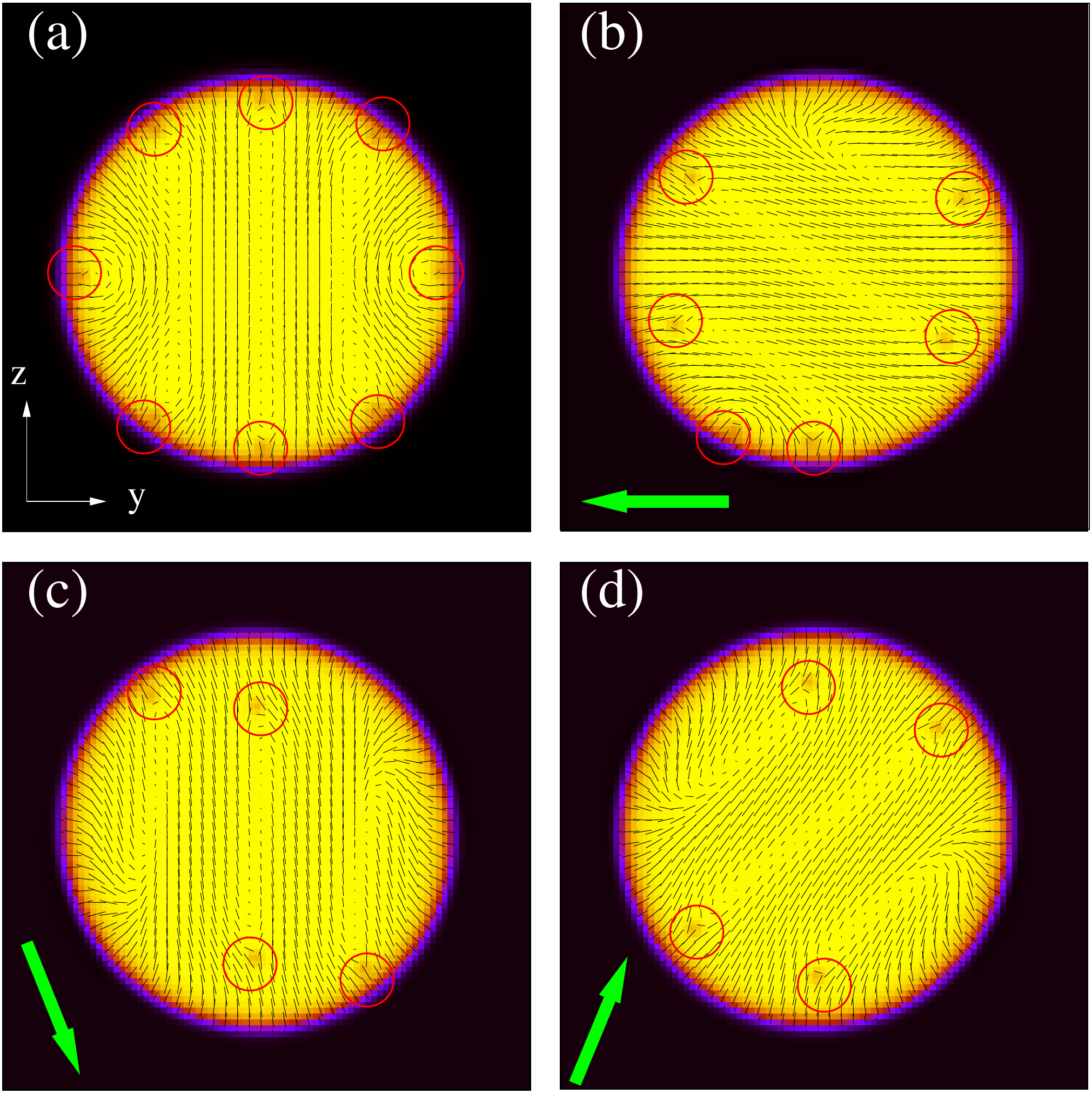}}
\caption{Dynamics of a cholesteric droplet (with $N=4$) under a counterclockwise rotating electric field. The reduced potential is with $\Delta V_y= \Delta V_z=4$, $\omega=10^{-3}$ and $W=-0.04$ (strong homeotropic anchoring). The motion is periodic and similar patterns appear each half turn of the director field. The green arrow indicates the direction of the electric field. Corresponding simulation times are 
$t=4\times 10^5$ (a), $t=4.5\times 10^5$ (b), $t=5.6\times 10^5$ (c), $t=7\times 10^5$ (d).}
\label{fig12}
\end{figure*}

If homeotropic anchoring is set on the surface, the equilibrium state has eight defects at the droplet surface (see Fig.~\ref{fig1}f) and three $+1$-charge regions in its bulk.
This structure leads to a more complex dynamics under a rotating electric field (Fig.~\ref{fig12}a-d and movie S2).
If it rotates counterclockwise, defects near the interface rotate coherently but, unlike the previous case, they remain  close to the interface and reduce first to six and finally to four.  Interestingly though, a twisted-like arrangement is still observed during a rotation; regions in which the director is almost fully aligned are spaced out by $+1$-charge regions with $\lambda$ defects, where the director escapes out of the plane. This structure appears periodically during the rotation, approximately after each half turn of the director field (see movie S2).
The velocity field pattern is overall similar to that of Fig.~\ref{fig11}, with $|{\bf u}|_{max}\simeq 2\times 10^{-4}$ (calculated in simulation time), almost half of the speed observed with tangential anchoring. This difference could be crudely explained in terms of a permeation-like effect. Indeed a vortex flow field, typically tangential nearby the interface, would encounter a larger resistance (namely an increased viscosity) if strong surface {\it homeotropic} anchoring is set rather than tangential due to the conflict bewtween the local direction of the velocity field and that of the director. 

\begin{figure*}[htbp]
\centerline{\includegraphics[width=1.1\textwidth]{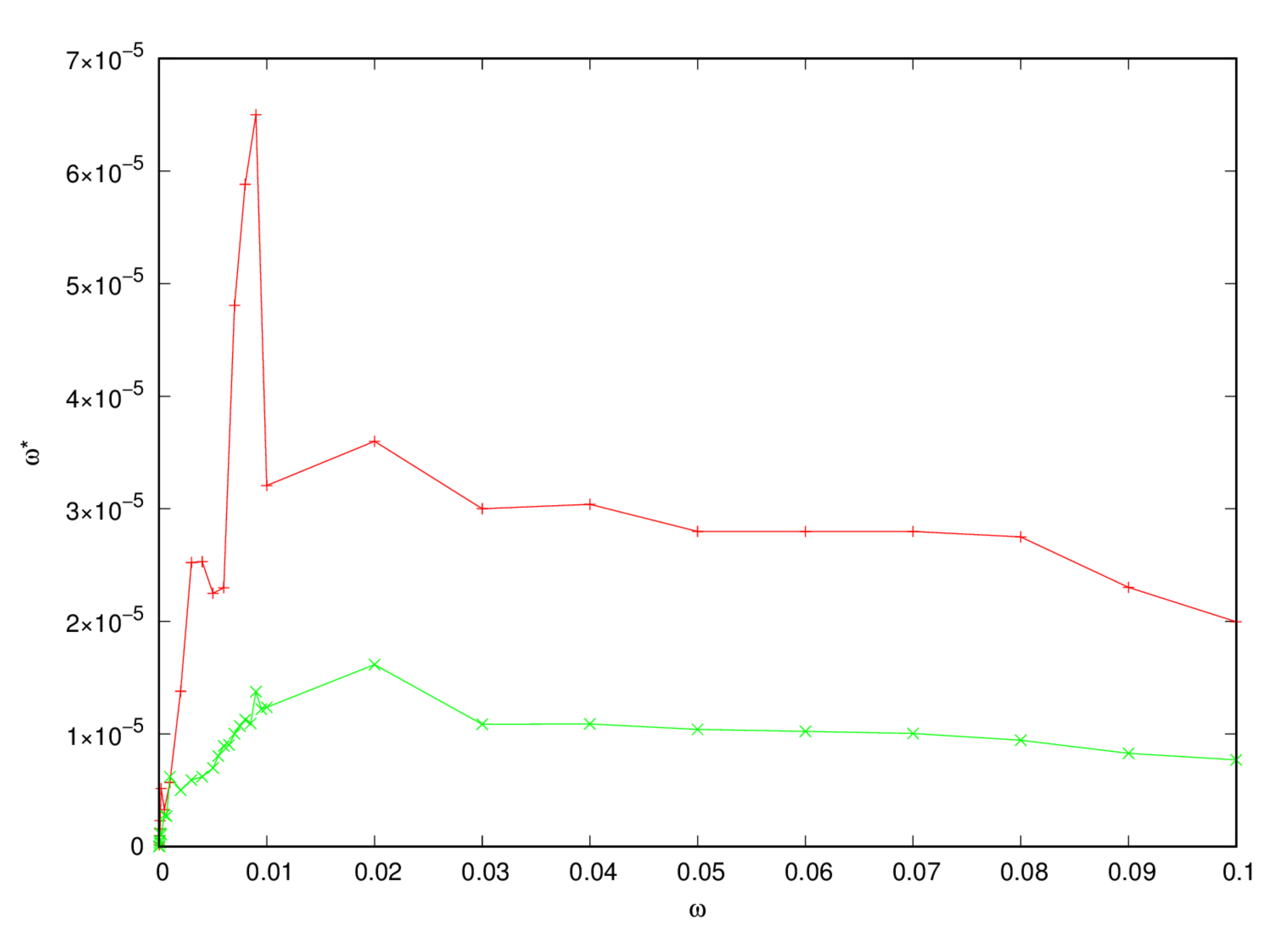}}
\caption{Angular velocity $\omega^*$ of a rotating cholesteric droplet calculated when $\omega$ ranges from $10^{-5}$ to $10^{-1}$ is shown for surface homeotropic anchoring (green/crosses) and surface tangential anchoring (red/plusses). A linear dependence is observed for $\omega\lesssim 10^{-2}$ in both cases, whereas for $\omega > 10^{-2}$ inertia dominates, and no rotation is observed.}
\label{fig15}
\end{figure*}

A signature of this behaviour might be seen by looking at how the angular velocity of the droplet $\omega^*$ (calculated as $\vec{\omega^*}=\frac{\int dV \phi {\bf r}\times {\bf u}}{\int dV r^2\phi}$) and the rotational velocity of the director field vary with the frequency  $\omega$ of the electric field. The two former quantities are crucial factors for controlling the defect dynamics and, more generally, for characterising the rheological response and the viscoelastic behaviour of the droplet.
We have looked at  the dynamics of the cholesteric droplet with both strong homogeneous and homeotropic surface anchoring for several values of $\omega$. In both cases we have found 
two regimes, one in which, for $\omega\lesssim 10^{-2}$, the angular velocity of the droplet  increases almost linearly, and another one in which, after a peak
achieved at $\omega\simeq 10^{-2}$,  the inertia of the droplet  dominates the dynamics and $\omega^*$ relaxes towards an almost constant value (see Fig.~\ref{fig15}).
When drag effects prevail, the droplet displays quick and short oscillations around a vertical axis passing through its centre of mass  and parallel to the $x$-direction. In this case  the velocity field patterns is closely reminiscent of that observed when an oscillating shear is applied to a droplet of liquid crystal~\cite{lamura}.
Note finally that the measured values of $\omega^*$ are red from two to three orders of magnitude smaller than those of $\omega$ (for values of $\omega\simeq 0.1$). 

In order to assess how the rotational speed difference between the director field and the electric field
affects the dynamics, a suitable quantity to compute is the Ericksen number $Er=\omega\Gamma R^2/K_{lc}$, a dimensionless number which describes the deformation of the director field under flow. 
If $\Gamma=1$, $R\sim 30$, $K_{lc}=0.03$ and $\omega\sim 10^{-3}$, we 
have $Er\sim 10$, meaning that our simulations are in a regime in which 
the viscous forces exceed the resistance due to the elasticity of the liquid crystal. 
Hence the director would be strongly advected by the fluid and no synchronization with the electric field would be observed.
Indeed we measure a rotational velocity of the director of $\simeq 2.3\times 10^{-6}$ for tangential surface anchoring  and $\simeq 1.05\times 10^{-6}$ for homeotropic surface anchoring,
values much lower than $\omega$. The different values observed in both cases could be possibly explained in terms of the permeation effect, more intense when homeotropic anchoring is considered.

\section{Conclusions}\label{conclusions}

We have studied, for the first time, the hydrodynamics of a cholesteric droplet in presence of  a (steady or rotating) field. In particular, by performing numerical  simulations, based on a lattice Boltzmann model of a cholesteric droplet in isotropic fluid, we have investigated its switching dynamics induced by an external electric field. We show that this process  is mainly affected by the surface anchoring, the magnitude and the direction of the electric field and the pitch of the cholesteric. An intriguingly  complex dynamics is usually observed when the field is applied parallel to the direction of the cholesteric axis and when $N=4$, as a higher number of $\pi$ twists creates more defects. 

 During switching, defects typically move from the interface towards the bulk of the emulsion where they can either get stuck or migrate back nearby the interface when the field is off, arresting the system into new metastable states. If the electric field is applied perpendicularly to the cholesteric axis the dynamics is overall simpler, as the director at equilibrium lies almost entirely along the direction of the electric field, hence its rotation and stretching is usually minimized. When the director is free to move at the droplet surface, the whole dynamics is defect-free and a switchable behaviour is found. The role of boudary conditions has also been investigated. Planar walls (with tangential anchoring), for instance, have been found to favour a nematic-like orientational order close to the walls, which gradually decreases when larger curvature (such as with circular domains) are considered. Besides the dynamic response under a uniform electric field, we have also studied the non-equilibrium effects, induced by a rotating field on the defect dynamics. While with surface tangential anchoring in-plane defects move towards the bulk of the droplet, with homeotropic anchoring their motion persists nearby the interface during the rotation and occurs with a defined periodicity. A key parameter to control the dynamics is the frequency $\omega$ of the electric field. Although the speed of rotation is usually one order of magnitude lower than $\omega$ (due to the anisotropy of liquid crystal), for low frequency it grows almost linearly, whereas for high frequency inertial effects dominate. 

The results are a first step towards a deeper understanding of the electric response of cholesteric emulsions and their possible uses in devices.  There are several directions to follow  for 
further  developments. For example, in several device-oriented applications it is crucial to assess the switching behaviour in response to a flexoelectric, rather than
a dielectric, field. Flexoelectricity is likely to be more important in materials in which the microscopic shape of molecules (e.g. pear or banana shaped) allows a coupling between electric field
and induced elastic distortions (splay or bend)~\cite{Coles}.  In this study we consider only a cholesteric liquid crystal made up of rod-like molecules, in which the flexoelectricity is negligible.
It would be also desirable to abandon the one elastic constant approximation and to perform a more systematic analysis of both the equilibrium  and switching dynamics of 
cholesteric emulsion where the three elastic energies (bend, twist and splay) can compete and richer defect patterns are expected.
Simulations in which several droplets are present in the system is also worth being investigated, either in a standard monodisperse or  polydisperse (i.e. cholesteric droplets with different helical pitches and surface anchoring) set up.  Finally, although the current code is fully three dimensional, obtaining results in three dimensions is very demanding in terms of computational resources but certainly of great interest for practical purposes. This is particularly true if one wants to simulate very large systems in order to minimize finite size effects and to reduce the artificial interactions
between the images of a droplet in a periodic lattice. On the other hand our quasi-2D system 
could be experimentally realized by considering, for instance, a thin film of liquid
crystal in which an emulsion is embedded. An example can be that described in Ref.~\cite{cluzeau}
in which a freely suspended smectic-C film with intrusions (namely droplets obtained by nucleation) is reported. Here droplets are observed for thin and intermediate sizes of the film.
A similar isotropic thin film in which cholesteric droplets are embedded would be a possible experimental realisation of our system.

\section*{Supplementary Material}

See supplemental material for movie S1 and movie S2. The first one describes the dynamics of a cholesteric droplet subject to a rotating electric field
when tangential anchoring is set on the interface. The second one describes the  dynamics of a cholesteric droplet subject to a rotating electric field
when homeotropic (or perpendicular) anchoring is set on the interface.

\section*{Acknowledgements}
Simulations ran on IBM Nextscale GALILEO at CINECA (Project INF16-fieldturb) under CINECA-INFN agreement and at Bari ReCaS e-Infrastructure funded by MIUR through
PON Research and Competitiveness 2007-2013 Call 254 Action I.

\bibliographystyle{unsrt}
\bibliography{biblio}% Produces the bibliography via BibTeX.

\begin{thebibliography}{10}
\bibitem{note1}
In each of these stretched $+1$-charge regions, there are two $\lambda^{+1/2}$
  defects at the edges.
\end{thebibliography}

\end{document}